\documentclass[a4paper,11pt]{article}
\usepackage{jcappub} 
\usepackage{lineno}
\usepackage[utf8]{inputenc}
\usepackage{amsmath,amssymb}
\usepackage{graphicx}
\usepackage{hyperref}
\usepackage{natbib}
\usepackage[normalem]{ulem}

\usepackage{multibib}
\usepackage[misc]{ifsym}
\usepackage{adjustbox}
\usepackage{multirow}
\usepackage{makecell}
\usepackage{natbib}

\allowdisplaybreaks
\title{Interacting Barrow Holographic Dark Energy in Non-flat Universe}
\author{Priyanka~Adhikary}
\author{and Sudipta~Das}
\affiliation{
Department of Physics, Visva-Bharati, Santiniketan 731235, India}
\emailAdd{sudipta.das@visva-bharati.ac.in}

\abstract{
Barrow holographic dark energy model is an extension of holographic dark energy that incorporates modifications to entropy due to quantum gravitational effects. In this work we study the cosmological properties of interacting Barrow holographic dark energy model in the case of non-zero curvature universe.
We construct the differential equations governing the evolution of the Barrow holographic dark energy density parameter and the dark matter density parameter in coupled form for both closed and open spatial geometry. Considering three different forms of coupling, we obtain the corresponding analytical expressions for the equation of state parameter for the dark energy component. We confront the scenario using recent observational datasets like cosmic chronometer and Pantheon data. It has been found that the strength of interaction as well as the curvature contribution come out to be nonzero which indicates that a non-flat interacting scenario is preferred by observational data.}

\begin{document}
\maketitle
\flushbottom
%%%%%%%%%%%%%%%%%%%%%%%%%%%%%%%%%%%%%%%%%%%%%%%%%%%%%%%%%%%%%%%%%%%%%%%%%%%%%%%%
\section{Introduction}
The evidence for cosmic acceleration is supported by observational data from Type Ia supernovae, CMB measurements, BAO, large-scale structure (LSS) data etc. \cite{riess1998observational, perlmutter1999measurements, arnaud2016planck, ahn2012ninth, jarosik2011seven}. Within the purview of general theory of relativity, this accelerated expansion is considered to be driven by dark energy (DE) component, which accounts for almost 70\% of the total energy density of the universe and should have a large negative pressure to drive the accelerated expansion \cite{padmanabhan2006dark}. However, the true nature of dark energy — whether it is a cosmological constant, a dynamical component or something else, remains one of the biggest open questions in cosmology. A number of cosmological models \cite{Copeland:2006wr,Cai:2009zp, sahni2006reconstructing, bamba2012dark, armendariz2001essentials, caldwell2002phantom, carroll2003can, kamenshchik2001alternative, sen2002tachyon, padmanabhan2002accelerated, copeland2005needed, amendola2010dark} 
have been proposed to resolve this issue considering modifications of matter or geometry sector of the Einstein equation, but no single theory can be definitively considered as the best candidate for dark energy.

The holographic dark energy (HDE) model \cite{Li:2004rb, Wang:2016och} is an interesting alternative approach to explain dark energy using the holographic principle, which arises from considerations in quantum gravity, black hole thermodynamics and string theory. The holographic principle suggests that all the information contained in a volume of space can be encoded on its boundary and the maximum entropy (information content) in a region of space grows with the surface area of the region  rather than its volume \cite{hooft2009, Susskind_1995}. Holographic dark energy models have been found to exhibit interesting cosmological aspects consistent with observations \cite{HorvatPRD2004, Pav_n_2005, Wang_2005, Nojiri_2006, SETARE2009331, MiaoLi_2009, AgostinoPRD2019, Molavi_2019}. The simplicity and reasonability of HDE provides a reliable framework to investigate the nature of dark energy. The energy density for holographic dark energy component depends on the cosmological length scale $L$. Different choices for the holographic cutoff length scale (such as the size of the cosmic horizon, the Hubble scale, or a particle horizon) lead to different interpretations of entropy in cosmological models. 

Recently Barrow proposed a black hole entropy relation that modifies the usual relationship between entropy and the area of the horizon by incorporating the idea that quantum gravitational effects could lead to a more complex, fractal structure of spacetime at small scales \cite{Barrow_2020}. The entropy-area relation for Barrow entropy is given by \cite{Barrow_2020}  
\begin{equation}
\label{Barrentropyy}
S_B=  \left (\frac{A}{A_0} \right )^{1+\frac{\Delta}{2}}, 
\end{equation}
with $A$ and $A_0$ being the standard horizon area and the Planck area respectively. The exponent $\Delta$, known as the Barrow exponent, accounts for the quantum gravitational deformation effects which lies in the range $0 \le \Delta \le 1$, such that $\Delta=0$ corresponds to the standard smooth structure and $\Delta=1$ corresponds to the most complex fractal universe. Application of this modified entropy relation yields a new class of holographic dark energy model known as Barrow holographic dark energy (BHDE) model \cite{Saridakis:2020PRD}. The BHDE model highlights the potential influence of quantum-gravitational effects on cosmic dynamics and suggests that small-scale quantum corrections can have a profound impact on large-scale cosmic phenomena. BHDE models have been shown to be a member of generalized holographic dark energy family \cite { Nojiri_2006, Nojiri_2017, Nojiri_2022, 2021NojiriSymmetry}. Cosmological dynamics of this class of models have been studied by a number of authors \cite{Saridakis:2020PRD,Anagnostopoulos:2020ctz,Saridakis:2020lrg, Mamon:2020spa, Barrow_2021,Das:2020rmg, Bhardwaj:2021chg, Chakraborty:2021uzp,Nojiri_2022_PLB, Nojiri:2022aof}. The evolution equations for the effective DE density parameter and the corresponding analytical expressions for both spatially flat and non-flat cases has been obtained in this modified cosmological scenario \cite{Saridakis:2020PRD, Adhikary:2021}. The thermodynamical properties of Barrow holographic dark energy (BHDE) models have been studied and the validity of the generalized second law has been explored in this context \cite{Mamon:2020spa, Chakraborty:2021uzp}. Attempts have been made to constrain the Barrow exponent $\Delta$ from various observational dataset including the EHT data from $M87^*$ and $S2$
star observations \cite{Luciano_2022, Jusufi:2021fek} and it has been found that the value of Barrow exponent is tightly constrained to very small values. BHDE models have been explored in modified gravity scenario as well. It has been shown that reconstructed $f(R)$, $f(Q,T)$ or scalar tensor gravity with BHDE provides an equation of state for the DE parameter which exhibits an evolutionary dynamics with the sequence of matter and dark energy epochs \cite{SarkarIJGMMP2021, Myrzakulov:2024qtd, Sobhanbabu:2024zmv}. These studies indicate that Barrow cosmology provides a new
background for studying various models of dark energy.

However, as the exact nature of these DE components is still unknown, it has been argued that an interaction between the dark matter and dark energy components could be a promising framework and can be useful in resolving the problems of standard model of cosmology, viz, the cosmological constant problem \cite{amendola2010dark, Yuri2015IJMPD}, Hubble tension etc. \cite{DiValentino:2017iww, Kumar:2019wfs}. Following this motivation, in the present work, we are interested in investigating the cosmological implications of Barrow Holographic dark energy in an interacting scenario for a non-flat universe. Interacting Barrow Holographic Dark energy models (IBHDE) have been studied by few authors considering various simplified choices for the holographic cut-off length scale, such as Hubble horizon cut-off \cite{Mamon:2020spa, Luciano:2022hhy}. But Hubble horizon is a more ``approximate" boundary compared to the future event horizon and is not an absolute or permanent boundary. Sheykhi et al. \cite{Sheykhi:2022fus} have studied interacting BHDE considering a future event horizon cut-off, but they have carried out the analysis in a spatially flat background. However the combined analysis of CMB anisotropy power spectra of the Planck Collaboration along with  the luminosity 
distance data indicates that a non-zero universe is favored at 99\%  confidence level \cite{DiValentino:2020hov}. Hence it would be interesting to investigate the properties of interacting BHDE model by considering future event horizon cut-off in a non-flat scenario. This consideration will provide the most general framework and might open some new possibilities regarding the true nature of dark sectors.

The paper is organized as follows: In section \ref{framework}  we present the basic equations for IBHDE model for both closed and open Friedmann-Robertson-Walker (FRW) metric assuming three different kinds of interacting terms. In section \ref{positivecase} and section \ref{negativecase}, we present the analytical expressions for various relevant cosmological parameters of the proposed model for different choices of the interaction term. In section \ref{Cosmological behavior} we present a detailed investigation of the cosmological 
behavior of the proposed interacting model for positive and negative spatial curvature cases respectively and observational constraints on various parameters of the model have been reported in section \ref{Observational Constraints}. Finally, we summarize our results in section \ref{conclusion}.

\section{Interacting Barrow holographic dark energy for non-zero spatial curvature}
\label{framework}

In this section we consider the background dynamics for a holographic dark energy model with non-zero spatial curvature taking into account an interaction between the holographic dark energy and the matter sectors. The Friedmann-Robertson-Walker (FRW) line element for the non-flat case is given by
\begin{equation}
 ds^2 = -\;dt^2 + a^2(t)\left[\frac{dr^2}{1-kr^2} +r^2 d\Omega^2 \right],
\end{equation}
where $a(t)$ is the scale factor and $k=+1,~0,~-1$ corresponds to closed, flat and open spatial curvature indices respectively.

The holographic dark energy density corresponding to Barrow entropy (\ref{Barrentropyy}) is obtained in  the form \cite{ Saridakis:2020PRD, Adhikary:2021} 
\begin{equation}\label{rhoDE}
 \rho_{DE} = C L^{\Delta-2},
\end{equation}
where  $L$ is the holographic horizon length and $C=3c^2 {M_P}^2$ is a  
parameter having dimensions  $[L]^{-2-\Delta}$ \cite{Adhikary:2021}. The expression for $\rho_{DE}$ given in equation (\ref{rhoDE}) can be obtained from the definition of
the standard holographic dark energy by imposing the condition that $S \propto A \propto L^2$ \cite{Wang:2016och}.
For $\Delta = 0$, the Barrow entropy reduces to the usual Bekenstein-Hawking entropy as $\rho_{DE} 
=C L^{-2}$ \cite{Li:2004rb,Wang:2016och}.\\  
In this framework, the Einstein's equations are written as
\begin{eqnarray}
3 H^2 +3\frac{k}{a^2} =\frac{1}{M_p^2} \left({\rho}_{m} + {\rho}_{DE}\right)\label{fe1}\\
2\dot{H} + 3 H^2 + \frac{k}{a^2} = -\frac{1}{M_p^2} p_{DE} \label{fe2},
\end{eqnarray}
where $H\equiv \frac{\dot{a}}{a}$ is the Hubble parameter, $\rho_{m}$, $\rho_{DE}$ are the energy densities corresponding to the matter sector and the Barrow holographic dark energy sectors respectively and $p_{DE}$ corresponds to the pressure component for the Barrow holographic dark energy sector. An overhead dot will indicate differentiation with respect to the cosmic time $t$. 

If the dark energy and dark matter sectors are considered to be non-interacting, then these two components will remain conserved by themselves. But as the true nature of these components are still unknown, an interacting framework may provide a more general scenario and may be instrumental in alleviating the cosmological coincidence problem which refers to the comparable energy densities of the dark energy and dark matter sectors at the present era even though they have completely different evolution dynamics \cite{Yuri2015IJMPD,zimdahl2001PLB, AP2000PRD, Salvatelli2014PRL,  mukherjee2017search,SD2014ASS,sinha2021prd}. 

We consider that only cold dark matter and
dark energy components interact with each other and contribute to the energy budget together which leads to the continuity equations for dark energy and matter sector as
\begin{eqnarray}
&&{\dot{\rho}}_{m} + 3 H \rho_{m} = Q\label{rhoconservm}\\
&&{\dot{\rho}}_{DE} + 3 H \left( 1+w_{DE}\right)\rho_{DE} = -Q\label{rhoconservde},
\end{eqnarray}
where $w_{DE}=\frac{p_{DE}}{\rho_{DE}}$ is the equation of state parameter (EOS) for the dark energy model. In most of the cosmological models, the source term $Q$ is chosen phenomenologically and the chosen form does not follow from the action principle. There has been attempts to derive the interaction term directly from the
action or the Lagrangian considering both classical as well as quantum field aspects of the dark energy component \cite{Wang_2016}, but it has been found that the coupling term cannot be determined properly and suffers
from hidden fine tuning problems \cite{Costa_2017}. For this reason the phenomenological choices are usually considered for studying interacting models, the most popular forms of the phenomenological interaction term being $Q = Q\left(H \rho_m \right)$ or $Q = Q\left(H \rho_{de} \right)$ or a combination of both. The reason for such a choice lies in its mathematical simplicity; because as evident from equations (\ref{rhoconservm}) and (\ref{rhoconservde}), the source term $Q$  must be a function of the energy densities multiplied by a quantity having units of $(\mathrm{time})^{-1}$ so as to keep the equations dimensionally consistent. In this context, as the Hubble parameter $H$ has the dimension of $(\mathrm{time})^{-1}$, this becomes a natural choice. For details regarding the cosmological implications of these phenomenological choices, one can look at \cite{Wang_2016}. Following the same arguments, in this work we consider three different choices for the interaction term $Q$ as:
\begin{eqnarray}
&&\mathrm{Case ~I ~~:~~} Q=-\Gamma H \rho_{DE} \label{case I}   \\
&&\mathrm{Case ~II ~~:~~}Q=-\Gamma H \rho_{m}=-\Gamma Hr \rho_{DE} \label{case II}  \\
&&\mathrm{Case ~III ~~:~~}Q=-\Gamma H \left(\rho_{m} + \rho_{DE} \right)=-\Gamma H(1+r) \rho_{DE} \label{case III}  
\label{threecases}
\end{eqnarray}
where $r=\frac{\rho_{m}}{\rho_{DE}}$ is the ratio of the two energy density components and $\Gamma$ is the strength of interaction. $\Gamma > 0$ or equivalently $Q < 0$ will indicate that that the flow of energy is from the matter sector to the DE sector whereas $\Gamma < 0$ will indicate the opposite scenario. In most of the interacting DM-DE models, the flow of energy is considered to be from the DM to the DE sector such that the dark energy sector becomes dominant at later times. However, Pavon and Wang \citep{Pav_n_2008} have shown that if dark energy is described to be a fluid with a temperature not far from equilibrium, the overall energy transfer should be from DE to DM sector for the second law of thermodynamics and Le Chatelier-Braun principle to remain valid.  Throughout this work we will consider $\Gamma$ to be positive so that the DE sector grows at the expense of the matter sector and dominates the evolution dynamics at later times. We have also introduced the standard observable parameters  like the effective equation-of-state parameter $w_{DE}\equiv\frac{p_{DE}}{\rho_{DE}}$ and the density parameters $\Omega_{m}\equiv\frac{\rho_m}{3M_p^2H^2}$, $\Omega_{DE}\equiv\frac{\rho_{DE}}{3M_p^2H^2}$ and $\Omega_{k}\equiv\frac{k}{a^2H^2}$ for the sake of convenience.

In order to study the evolution dynamics for a holographic dark energy model, one needs to suitably choose the horizon length $L$ appearing in equation (\ref{rhoDE}). In case of flat spatial geometry, there are many possible choices for the horizon length $L$, the most common one being the future event horizon $R_h$ \cite{Li:2004rb} given by
\begin{equation}
\label{futurehor}
R_h\equiv a\int_t^\infty \frac{dt}{a}= a\int_a^\infty \frac{da}{Ha^2}
\end{equation}
However in the case of non-flat spatial geometry, the above length should be modified  suitably \cite{Adhikary:2021,Huang_2004,SETARE20061}. The corresponding modifications will be different for closed $(k=+1)$ and open $(k=-1)$ cases and we will consider them separately in the following sections.
     
\section {Cosmological behavior of interacting models with positive spatial curvature} \label{positivecase}
For a closed universe ($k=+1$), the horizon length $L$ is given by  $L= a~ r_h(t)$, where $r_h(t)$ is determined from the relation \cite{ Adhikary:2021, Huang_2004,SETARE20061} 
\begin{equation}
\int_{0}^{r_h(t)} \frac{dr'}{\sqrt{1- {r'}^2}}= 
\frac{R_h}{a }. 
\end{equation}
This gives,
\begin{equation}
\label{r}
%r(t) = \frac{1}{\sqrt{k}}\sin y , 
r_h(t) = \sin y, 
\end{equation}
 where  
\begin{equation}
\label{ydefnit}
%y=\sqrt{k} \frac{R_h}{a}  = \sqrt{k} 
%\int_{x}^{\infty} \frac{dx}{a H},
y=\frac{R_h}{a}  = \int_{x}^{\infty} \frac{dx}{a H}, 
\end{equation}
and $x=\ln a$.\\
Now equation (\ref{fe1}) can be rewritten as
\begin{equation}
\label{kequation}
\frac{1}{a^2H^2}= \Omega_m + \Omega_{DE}-1 
\end{equation}
Using $\Omega_{DE} = \frac{\rho_{DE}}{3M_p^2H^2}$, $\Omega_{k0} = \frac{1}{a_0^2 H_0^2}$ and $\rho_{DE} = 3c^2 {M_P}^2 L^{\Delta-2}$ from (\ref{rhoDE}), one gets
 \begin{equation}\label{L}
 L=\left[\frac{\Omega_{DE}\left(\Omega_{k0}H_0^2\right)(1+z)^{2}}{c^2(\Omega_{DE}+\Omega_{m}-1)}\right]^\frac{1}{\Delta-2}
\end{equation}
Hence, from equations (\ref{r}), (\ref{ydefnit}) and (\ref{L}) along with $L= a~ r_h(t)$, we obtain
\begin{equation}\label{siny}
 \sin{\left[\int_{x}^{\infty} \frac{dx}{a H}\right]}=\frac{1}{a}\left[\frac{\Omega_{DE}\left(\Omega_{k0}H_0^2\right)(1+z)^{2}}{c^2(\Omega_{DE}+\Omega_{m}-1)}\right]^\frac{1}{\Delta-2}
\end{equation}
Now for different forms of interaction $Q$ considered in equations (\ref{case I}) to (\ref{case III}), $\Omega_{DE}$ and $\Omega_{m}$ will depict different evolution history of the universe. In the next subsections, we will consider the cosmological analysis for the non-flat Barrow holographic dark energy model considering the three different forms of interaction term $Q$ mentioned earlier. 
%%%%%%%%%%%%%%%%%%%%%%%%%%%%%%%%%%%%%%%%
\subsection{Case I: $Q=-\Gamma H \rho_{DE}$}\label{modelI}
With $Q=-\Gamma H \rho_{DE}$, equation (\ref{rhoconservde}) can be integrated to obtain 
\begin{equation}\label{lnrho}
 \mathrm{ln} \left(\frac{\rho_{DE}}{\rho_{0}}\right)=\int{\left[\Gamma H -3H(1+w_{DE})\right]} dt
\end{equation}
$\rho_0$ being the constant of integration.\\ 
Differentiating equation (\ref{lnrho}) with respect to $x=\ln a$ and using the relation $\frac{dt}{dx}=\frac{1}{H}$, one obtains
\begin{equation}\label{lnrhodx}
\frac{d}{dx}\left[\mathrm{ln} \left(\frac{\rho_{DE}}{\rho_{0}}\right)\right]=\Gamma -3(1+w_{DE})
\end{equation}
Now the left hand side of equation (\ref{lnrhodx}) can be expressed in terms of $\Omega_{DE}$ as
\begin{equation}\label{lnrhoomega}
\frac{d}{dx}\left[\mathrm{ln} \left(\frac{\rho_{DE}}{\rho_{0}}\right)\right]= \frac{1}{\rho_{DE}}\frac{d}{dx}\left[\frac{\Omega_{DE}}{(1-\Omega_{DE})}\left(\rho_m - \frac{\beta}{a^2}\right)\right]
\end{equation}
where $\beta=3 M_p^2$. Equation (\ref{lnrhoomega}) can be further simplified to obtain
\begin{equation}\label{lnrhoomega1}
\frac{d}{dx}\left[\mathrm{ln} \left(\frac{\rho_{DE}}{\rho_{0}}\right)\right]=\frac{\Omega_{DE}'}{\Omega_{DE}(1-\Omega_{DE})} + \frac{1}{\left(\rho_m -\frac{\beta}{a^2}\right)} \frac{d}{dx}\left(\rho_m -\frac{\beta}{a^2}\right)
\end{equation}
Again equation (\ref{rhoconservm}) with $Q=-\Gamma H \rho_{DE}$ gives
\begin{equation}\label{g1}
\frac{1}{\rho_m}\frac{d \rho_m}{dx} = -\left(\frac{\Gamma}{r} + 3\right)
\end{equation}
where $r=\frac{\rho_{m}}{\rho_{DE}}=\frac{\Omega_{m}}{\Omega_{DE}}$. This immediately gives 
\begin{equation}\label{lnrhoomega2}
\frac{1}{\left(\rho_m -\frac{\beta}{a^2}\right)} \frac{d}{dx}\left(\rho_m -\frac{\beta}{a^2}\right)=\frac{1}{(1-\Omega_{DE})} \left[ 2(\Omega_{m}+\Omega_{DE}-1)-\Omega_{m}\left(\frac{\Gamma}{r}+3\right)\right]
\end{equation}
Using equations (\ref{lnrhodx}) and   (\ref{lnrhoomega2}), equation (\ref{lnrhoomega1}) becomes
\begin{equation}\label{lnrhoomega3}
\frac{\Omega_{DE}'}{\Omega_{DE}(1-\Omega_{DE})} + \frac{1}{(1-\Omega_{DE})} \left[ 2(\Omega_{m}+\Omega_{DE}-1)-\Omega_{m}\left(\frac{\Gamma}{r}+3\right)\right]=\Gamma -3(1+w_{DE})
\end{equation}
Again time derivative of $\rho_{DE} = 3c^2 {M_P}^2 L^{\Delta-2}$ gives
\begin{equation}
{\dot{\rho}}_{DE} = (\Delta-2) H \left(1-\frac{\cos y}{HL}\right)\rho_{DE}
\end{equation}
where $L=a r_h(t)=a \sin y$ for $k=+1$ model. Putting this expression in equation (\ref{rhoconservde}), one obtains the expression for $w_{DE}$ for the interaction form considered in case I as
\begin{equation}\label{model1wde}
w_{DE} =- 
\left(\frac{1+\Delta}{3}\right)+\frac{\Delta-2}{3}\left(\cos y \sqrt{\frac{3 M_p^2 \Omega_{DE}}{C}}  L^{-\frac{\Delta}{2}}\right)+\frac{\Gamma}{3}.
\end{equation}
Using (\ref{model1wde}) in (\ref{lnrhoomega3}), one obtains the expression for evolution of interacting Barrow holographic dark energy model in a closed
universe as
\begin{equation}\label{model1de}
\begin{split}
\frac{\Omega_{DE}'}{\Omega_{DE} (1-\Omega_{DE})} = (\Delta - 2)\left(1-\cos y \sqrt{\frac{3 M_p^2 \Omega_{DE}}{C}} L^{-\frac{\Delta}{2}}\right)\\ - \frac{1}{(1-\Omega_{DE})} \left[ 2(\Omega_{m}+\Omega_{DE}-1)-\Omega_{m}\left(\frac{\Gamma}{r}+3\right)\right]
\end{split}
\end{equation}
In a similar way starting with (\ref{rhoconservm}) and using the expression for the source term given in (\ref{case I}), we acquire 
\begin{equation}\label{model1m}
\begin{split}
\frac{\Omega_{m}'}{\Omega_{m}} =-\left(\frac{\Gamma}{r}+3\right)-\Omega_{DE} (\Delta - 2)\left(1-\cos y \sqrt{\frac{3 M_p^2 \Omega_{DE}}{C}} L^{-\frac{\Delta}{2}}\right)\\ - \left[ 2(\Omega_{m}+\Omega_{DE}-1)-\Omega_{m}\left(\frac{\Gamma}{r}+3\right)\right]
\end{split}
\end{equation}
Throughout the text, a prime will indicate differentiation with respect to  $x=\ln a$. Detailed calculations for section \ref{modelI} has been provided in appendix \ref{appendix1}.\\
Differential equations (\ref{model1de}) and (\ref{model1m}) will govern the evolution of Barrow holographic dark energy and matter sector respectively in a closed universe for this particular choice of $Q$. In the case where $\Gamma=0$, it provides the evolution equations for $\Omega_{DE}$ and $\Omega_{m}$ corresponding to a Barrow holographic dark energy model in non-flat universe \cite{Adhikary:2021, Huang_2004}. Additionally, in the case where $\Delta=0$ and $\Gamma=0$, it coincides with the usual holographic dark energy model in a closed universe \cite{Saridakis:2020PRD, SETARE20061}.
%%%%%%%%%%%%%%%%%%%%%%%%%%%%%%%%%%%%%%%%%%%%
\subsection{$\mathrm{Case ~II ~~:~~} Q=-\Gamma H \rho_{m}=-\Gamma H r\rho_{DE}$}
With $Q=-\Gamma H r \rho_{DE}$, equation (\ref{rhoconservde}) can be integrated to obtain 
\begin{equation}\label{lnrho2}
 \mathrm{ln} \left(\frac{\rho_{DE}}{\rho_{0}}\right)=\int{\left[\Gamma H r -3H(1+w_{DE})\right]} dt
\end{equation}
As done in the previous section, differentiation of equation (\ref{lnrho2}) with respect to $x=\ln a$ gives
\begin{equation}\label{lnrhodx2}
\frac{d}{dx}\left[\mathrm{ln} \left(\frac{\rho_{DE}}{\rho_{0}}\right)\right]=\Gamma r -3(1+w_{DE})
\end{equation}
Again equation (\ref{rhoconservm}), for this particular form of $Q=-\Gamma H r\rho_{DE}$, gives
\begin{equation}
\frac{1}{\rho_m}\frac{d \rho_m}{dx} = -\left(\Gamma + 3\right)
\end{equation}
Following the same calculations as carried out in section \ref{modelI}, one finally arrives at
\begin{equation}\label{lnrhoomega3r2}
\frac{\Omega_{DE}'}{\Omega_{DE}(1-\Omega_{DE})} + \frac{1}{(1-\Omega_{DE})} \left[ 2(\Omega_{m}+\Omega_{DE}-1)-\Omega_{m}\left(\Gamma+3\right)\right]=\Gamma r -3(1+w_{DE})
\end{equation}
Again taking the time derivative of $\rho_{DE}$ and putting this expression in equation (\ref{rhoconservde}) along with $Q=-\Gamma H r \rho_{DE}$, one obtains the expression for $w_{DE}$ for case II as
\begin{equation}\label{model2wde}
w_{DE} =- 
\left(\frac{1+\Delta}{3}\right)+\frac{\Delta-2}{3}\left(\cos y \sqrt{\frac{3 M_p^2 \Omega_{DE}}{C}}  L^{-\frac{\Delta}{2}}\right)+\frac{\Gamma r}{3}.
\end{equation}
Using (\ref{model2wde}) in equation(\ref{lnrhoomega3r2}), one obtains the expression for evolution of interacting Barrow holographic dark energy model (using $Q=-\Gamma H r \rho_{DE}$) in a closed
universe as
\begin{equation}\label{model2de}
\begin{split}
\frac{\Omega_{DE}'}{\Omega_{DE} (1-\Omega_{DE})} = (\Delta - 2)\left(1-\cos y \sqrt{\frac{3 M_p^2 \Omega_{DE}}{C}} L^{-\frac{\Delta}{2}}\right)\\ - \frac{1}{(1-\Omega_{DE})} \left[ 2(\Omega_{m}+\Omega_{DE}-1)-\Omega_{m}\left(\Gamma+3\right)\right]
\end{split}
\end{equation}
In a similar way starting with (\ref{rhoconservm}) and using the expression for the source term given in (\ref{case II}), we acquire 
\begin{equation}\label{model2m}
\begin{split}
\frac{\Omega_{m}'}{\Omega_{m}} =-\left(\Gamma+3\right)-\Omega_{DE} (\Delta - 2)\left(1-\cos y \sqrt{\frac{3 M_p^2 \Omega_{DE}}{C}} L^{-\frac{\Delta}{2}}\right)\\ - \left[ 2(\Omega_{m}+\Omega_{DE}-1)-\Omega_{m}\left(\Gamma+3\right)\right]
\end{split}
\end{equation}
Differential equations (\ref{model2de}) and (\ref{model2m}) will govern the evolution of Barrow holographic dark energy and matter sector respectively in a closed universe with the source term being $Q=-\Gamma H \rho_{m}$. Again $\Delta=0$ and $\Gamma=0$ will provide the usual holographic dark energy model in a closed universe \cite{Saridakis:2020PRD, SETARE20061}.
%%%%%%%%%%%%%%%%%%%%%%%%%%%%%%%%%%%%%%%%%%
\subsection{$\mathrm{Case ~III ~~:~~} Q=-\Gamma H \left(\rho_{m}+\rho_{DE}\right)=-\Gamma H (1+r)\rho_{DE}$}
With $Q=-\Gamma H (1+r) \rho_{DE}$, equation (\ref{rhoconservde}) can be integrated to obtain 
\begin{equation}\label{lnrho3}
 \mathrm{ln} \left(\frac{\rho_{DE}}{\rho_{0}}\right)=\int{\left[\Gamma H (1+r) -3H(1+w_{DE})\right]} dt
\end{equation}
which with equation (\ref{lnrho3}) will give
\begin{equation}\label{lnrhodx3}
\frac{d}{dx}\left[\mathrm{ln} \left(\frac{\rho_{DE}}{\rho_{0}}\right)\right]=\Gamma (1+r) -3(1+w_{DE})
\end{equation}
Further, equation (\ref{rhoconservm}) with $Q=-\Gamma H (1+r)\rho_{DE}$ gives
\begin{equation}
\frac{1}{\rho_m}\frac{d \rho_m}{dx} = -\left( \frac{\Gamma (1+r)}{r} + 3\right)
\end{equation}
This immediately gives 
\begin{equation}\label{lnrhoomega2r3}
\frac{d}{dx}\left[\mathrm{ln} \left(\frac{\rho_{DE}}{\rho_{0}}\right)\right]=\frac{\Omega_{DE}'}{\Omega_{DE}(1-\Omega_{DE})} + \frac{1}{(1-\Omega_{DE})} \\ \left[ 2(\Omega_{m}+\Omega_{DE}-1)-\Omega_{m}\left(\frac{\Gamma (1+r)}{r} + 3\right)\right]
\end{equation}
Using equations (\ref{lnrhodx3}) and   (\ref{lnrhoomega2r3}), one gets
\begin{equation}\label{lnrhoomega3r3}
\begin{split}
\frac{\Omega_{DE}'}{\Omega_{DE}(1-\Omega_{DE})} + \frac{1}{(1-\Omega_{DE})}  \left[ 2(\Omega_{m}+\Omega_{DE}-1)-\Omega_{m}\left(\frac{\Gamma (1+r)}{r} + 3\right)\right]\\
=\Gamma (1+r) -3(1+w_{DE})
\end{split}
\end{equation}
Using the time derivative of $\rho_{DE}$ 
and putting this expression in equation (\ref{rhoconservde}), one obtains the expression for $w_{DE}$ for case III as
\begin{equation}\label{model3wde}
w_{DE} =- 
\left(\frac{1+\Delta}{3}\right)+\frac{\Delta-2}{3}\left(\cos y \sqrt{\frac{3 M_p^2 \Omega_{DE}}{C}}  L^{-\frac{\Delta}{2}}\right)+\frac{\Gamma (1+r)}{3}.
\end{equation}
Using (\ref{model3wde}) in equation(\ref{lnrhoomega3r3}), one obtains the expression for evolution of interacting Barrow holographic dark energy model (using $Q=-\Gamma H (1+r)\rho_{DE}$) in a closed
universe as
\begin{equation}\label{model3de}
\begin{split}
\frac{\Omega_{DE}'}{\Omega_{DE} (1-\Omega_{DE})} = (\Delta - 2)\left(1-\cos y \sqrt{\frac{3 M_p^2 \Omega_{DE}}{C}} L^{-\frac{\Delta}{2}}\right)\\ - \frac{1}{(1-\Omega_{DE})} \left[ 2(\Omega_{m}+\Omega_{DE}-1)-\Omega_{m}\left(\frac{\Gamma (1+r)}{r}+3\right)\right]
\end{split}
\end{equation}
In a similar way starting with (\ref{rhoconservm}) and using the expression for the source term given in (\ref{case III}), we obtain 
\begin{equation}\label{model3m}
\begin{split}
\frac{\Omega_{m}'}{\Omega_{m}} =-\left(\frac{\Gamma (1+r)}{r}+3\right)-\Omega_{DE} (\Delta - 2)\left(1-\cos y \sqrt{\frac{3 M_p^2 \Omega_{DE}}{C}} L^{-\frac{\Delta}{2}}\right)\\ - \left[ 2(\Omega_{m}+\Omega_{DE}-1)-\Omega_{m}\left(\frac{\Gamma (1+r)}{r}+3\right)\right]
\end{split}
\end{equation}
Differential equations (\ref{model3de}) and (\ref{model3m}) will govern the evolution of Barrow holographic dark energy model in a closed universe for the particular choice of $Q$ given in (\ref{case III}). 

%%%%%%%%%%%%%%%%%%%%%%%%%%%%%%%%%%%%%%%%%%%%%%%%%%%
\section {Cosmological behavior of interacting models with negative spatial curvature}\label{negativecase}
For an open universe ($k=-1$) model, $r_h(t)$ is determined from the relation \cite{ Adhikary:2021, Huang_2004,SETARE20061} 
\begin{equation}
\int_{0}^{r_h(t)} \frac{dr'}{\sqrt{1+ {r'}^2}}= 
\frac{R_h}{a }. 
\end{equation}
This gives,
\begin{equation}
\label{rn}
r_h(t) = \sinh y , 
\end{equation}
 where  
\begin{equation}
\label{ydefnitn}
y=\frac{R_h}{a}  = \int_{x}^{\infty} \frac{dx}{a H}, 
\end{equation}
Now for $k=-1$, equation (\ref{fe1}) can be rewritten as
\begin{equation}
\label{kequationn}
\frac{1}{a^2H^2}= 1- \Omega_m - \Omega_{DE} 
\end{equation}
Using $\Omega_{DE} = \frac{\rho_{DE}}{3M_p^2H^2}$, $\Omega_{k0} = -\frac{1}{a_0^2 H_0^2}$ and $\rho_{DE} = 3c^2 {M_P}^2 L^{\Delta-2}$ from (\ref{rhoDE}), one gets ($C=3c^2{M_P}^2$)
 \begin{equation}\label{Ln}
 L=\left[\frac{\Omega_{DE}(\Omega_{k0}H_0^2)(1+z)^2}{c^2(\Omega_{DE} + \Omega_{m}-1)}\right]^\frac{1}{\Delta-2}
\end{equation}
Although equations (\ref{L}) and (\ref{Ln}) have similar expressions, in equation (\ref{L}), $\Omega_{k0}$ is positive whereas in equation (\ref{Ln}), $\Omega_{k0}$ is negative.\\
From equations (\ref{rn}), (\ref{ydefnitn}) and (\ref{Ln}) along with $L= a~ r_h(t)$, we obtain
\begin{equation}\label{sinhy}
 \sinh{\left[\int_{x}^{\infty} \frac{dx}{a H}\right]}=\frac{1}{a}\left[\frac{\Omega_{DE}(\Omega_{k0}H_0^2)(1+z)^2}{c^2(\Omega_{DE} + \Omega_{m}-1)}\right]^\frac{1}{\Delta-2}
\end{equation}
Again considering the same forms of the interaction term $Q$ given in (\ref{case I}) to (\ref{case III}), the cosmological evolution for the negative curvature case has been studied in the next subsections. Detailed calculations for section \ref{modelI} and section \ref{modelnI} have been provided in appendix \ref{appendix1} and appendix \ref{appendix2}.
%%%%%%%%%%%%%%%%%%%%%%%%%%%%%%%%%%%%%%%%%
\subsection{$\mathrm{Case ~I ~~:~~} Q=-\Gamma H \rho_{DE}$}\label{modelnI}
For $Q=-\Gamma H \rho_{DE}$, one has (see equation (\ref{lnrhodx}) of section \ref{modelI})
\begin{equation}\label{lnrhodxn}
\frac{d}{dx}\left[\mathrm{ln} \left(\frac{\rho_{DE}}{\rho_{0}}\right)\right]=\Gamma -3(1+w_{DE})
\end{equation}
As before, the left hand side of equation (\ref{lnrhodxn}) can be expressed as
\begin{equation}\label{lnrhoomegan}
\frac{d}{dx}\left[\mathrm{ln} \left(\frac{\rho_{DE}}{\rho_{0}}\right)\right]= \frac{1}{\rho_{DE}}\frac{d}{dx}\left[\frac{\Omega_{DE}}{(1-\Omega_{DE})}\left(\rho_m + \frac{\beta}{a^2}\right)\right]
\end{equation}
where $\beta=3 M_p^2$ as before. This can be further simplified to obtain
\begin{equation}\label{lnrhoomega1n}
\frac{d}{dx}\left[\mathrm{ln} \left(\frac{\rho_{DE}}{\rho_{0}}\right)\right]=\frac{\Omega_{DE}'}{\Omega_{DE}(1-\Omega_{DE})} + \frac{1}{\left(\rho_m +\frac{\beta}{a^2}\right)} \frac{d}{dx}\left(\rho_m +\frac{\beta}{a^2}\right)
\end{equation}
Using equation (\ref{g1}) and following the same procedure as section \ref{modelI}, one obtains for $k=-1$ case
\begin{equation}\label{lnrhoomega2n}
\frac{1}{\left(\rho_m +\frac{\beta}{a^2}\right)} \frac{d}{dx}\left(\rho_m +\frac{\beta}{a^2}\right)=\frac{1}{(1-\Omega_{DE})} \left[ 2(\Omega_{m}+\Omega_{DE}-1)-\Omega_{m}\left(\frac{\Gamma}{r}+3\right)\right]
\end{equation}
Using equations (\ref{lnrhodxn}) and   (\ref{lnrhoomega2n}), equation (\ref{lnrhoomega1n}) becomes
\begin{equation}\label{lnrhoomega3n}
\frac{\Omega_{DE}'}{\Omega_{DE}(1-\Omega_{DE})} + \frac{1}{(1-\Omega_{DE})} \left[ 2(\Omega_{m}+\Omega_{DE}-1)-\Omega_{m}\left(\frac{\Gamma}{r}+3\right)\right]=\Gamma -3(1+w_{DE})
\end{equation}
Again time derivative of $\rho_{DE} = 3c^2 {M_P}^2 L^{\Delta-2}$ gives
\begin{equation}
{\dot{\rho}}_{DE} = (\Delta-2) H \left(1-\frac{\cosh y}{HL}\right)\rho_{DE}
\end{equation}
where $L=a r_h(t)=a \sinh y$ for $k=-1$ model. Putting this expression in equation (\ref{rhoconservde}), one obtains the expression for $w_{DE}$ as
\begin{equation}\label{model1wden}
w_{DE} =- 
\left(\frac{1+\Delta}{3}\right)+\frac{\Delta-2}{3}\left(\cosh y \sqrt{\frac{3 M_p^2 \Omega_{DE}}{C}}  L^{-\frac{\Delta}{2}}\right)+\frac{\Gamma}{3}.
\end{equation}
Using (\ref{model1wden}) in (\ref{lnrhoomega3n}), one obtains the expression for evolution of interacting Barrow holographic dark energy model in an open
universe as
\begin{equation}\label{model1den}
\begin{split}
\frac{\Omega_{DE}'}{\Omega_{DE} (1-\Omega_{DE})} = (\Delta - 2)\left(1-\cosh y \sqrt{\frac{3 M_p^2 \Omega_{DE}}{C}} L^{-\frac{\Delta}{2}}\right)\\ - \frac{1}{(1-\Omega_{DE})} \left[ 2(\Omega_{m}+\Omega_{DE}-1)-\Omega_{m}\left(\frac{\Gamma}{r}+3\right)\right]
\end{split}
\end{equation}
In a similar way starting with (\ref{rhoconservm}) and using the expression for the source term given in (\ref{case I}), we acquire 
\begin{equation}\label{model1mn}
\begin{split}
\frac{\Omega_{m}'}{\Omega_{m}} =-\left(\frac{\Gamma}{r}+3\right)-\Omega_{DE} (\Delta - 2)\left(1-\cosh y \sqrt{\frac{3 M_p^2 \Omega_{DE}}{C}} L^{-\frac{\Delta}{2}}\right)\\ - \left[ 2(\Omega_{m}+\Omega_{DE}-1)-\Omega_{m}\left(\frac{\Gamma}{r}+3\right)\right]
\end{split}
\end{equation}
Throughout the text, a prime will indicate differentiation with respect to  $x=\ln a$. 
Differential equations (\ref{model1den}) and (\ref{model1mn}) will govern the evolution of Barrow holographic dark energy and matter sector respectively in an open universe for this particular choice of $Q$. In the case where $\Gamma=0$, it provides the evolution equations for $\Omega_{DE}$ and $\Omega_{m}$ corresponding to a Barrow holographic dark energy model in non-flat universe \cite{Adhikary:2021, Huang_2004}. Additionally, in the case where $\Delta=0$ and $\Gamma=0$, it coincides with the usual holographic dark energy model in an open universe \cite{Saridakis:2020PRD, SETARE20061}. 
%%%%%%%%%%%%%%%%%%%%%%%%%%%%%%%%%%%%%%%%%%%%
\subsection{$\mathrm{Case ~II ~~:~~} Q=-\Gamma H \rho_{m}=-\Gamma H r\rho_{DE}$}
With $Q=-\Gamma H r \rho_{DE}$, equation (\ref{rhoconservde}) can be integrated to obtain 
\begin{equation}\label{lnrho2n}
 \mathrm{ln} \left(\frac{\rho_{DE}}{\rho_{0}}\right)=\int{\left[\Gamma H r -3H(1+w_{DE})\right]} dt
\end{equation}
As done in the previous section, differentiation of equation (\ref{lnrho2n}) with respect to $x=\ln a$ gives
\begin{equation}\label{lnrhodx2n}
\frac{d}{dx}\left[\mathrm{ln} \left(\frac{\rho_{DE}}{\rho_{0}}\right)\right]=\Gamma r -3(1+w_{DE})
\end{equation}
Again equation (\ref{rhoconservm}), for this particular form of $Q=-\Gamma H r\rho_{DE}$, gives
\begin{equation}
\frac{1}{\rho_m}\frac{d \rho_m}{dx} = -\left(\Gamma + 3\right)
\end{equation}
Following the same calculations as carried out in section \ref{modelI}, one finally arrives at
\begin{equation}\label{lnrhoomega3r2n}
\frac{\Omega_{DE}'}{\Omega_{DE}(1-\Omega_{DE})} + \frac{1}{(1-\Omega_{DE})} \left[ 2(\Omega_{m}+\Omega_{DE}-1)-\Omega_{m}\left(\Gamma+3\right)\right]=\Gamma r -3(1+w_{DE})
\end{equation}
Again taking the time derivative of $\rho_{DE}$ and putting this expression in equation (\ref{rhoconservde}) along with $Q=-\Gamma H r \rho_{DE}$, one obtains the expression for $w_{DE}$ as
\begin{equation}\label{model2wden}
w_{DE} =- 
\left(\frac{1+\Delta}{3}\right)+\frac{\Delta-2}{3}\left(\cosh y \sqrt{\frac{3 M_p^2 \Omega_{DE}}{C}}  L^{-\frac{\Delta}{2}}\right)+\frac{\Gamma r}{3}.
\end{equation}
Using (\ref{model2wden}) in equation(\ref{lnrhoomega3r2n}), one obtains the expression for evolution of interacting Barrow holographic dark energy model (using $Q=-\Gamma H r \rho_{DE}$) in an open universe as
\begin{equation}\label{model2den}
\begin{split}
\frac{\Omega_{DE}'}{\Omega_{DE} (1-\Omega_{DE})} = (\Delta - 2)\left(1-\cosh y \sqrt{\frac{3 M_p^2 \Omega_{DE}}{C}} L^{-\frac{\Delta}{2}}\right)\\ - \frac{1}{(1-\Omega_{DE})} \left[ 2(\Omega_{m}+\Omega_{DE}-1)-\Omega_{m}\left(\Gamma+3\right)\right]
\end{split}
\end{equation}
In a similar way starting with (\ref{rhoconservm}) and using the expression for the source term given in (\ref{case II}), we acquire 
\begin{equation}\label{model2mn}
\begin{split}
\frac{\Omega_{m}'}{\Omega_{m}} =-\left(\Gamma+3\right)-\Omega_{DE} (\Delta - 2)\left(1-\cosh y \sqrt{\frac{3 M_p^2 \Omega_{DE}}{C}} L^{-\frac{\Delta}{2}}\right)\\ - \left[ 2(\Omega_{m}+\Omega_{DE}-1)-\Omega_{m}\left(\Gamma+3\right)\right]
\end{split}
\end{equation}
Differential equations (\ref{model2den}) and (\ref{model2mn}) will govern the evolution of Barrow holographic dark energy and matter sector respectively in a open universe with the source term being $Q=-\Gamma H \rho_{m}$. Again $\Delta=0$ and $\Gamma=0$ will provide the usual holographic dark energy model in an open universe \cite{Saridakis:2020PRD, SETARE20061}.
%%%%%%%%%%%%%%%%%%%%%%%%%%%%%%%%%%%%%%%%%%
\subsection{$\mathrm{Case ~III ~~:~~} Q=-\Gamma H \left(\rho_{m}+\rho_{DE}\right)=-\Gamma H (1+r)\rho_{DE}$}
With $Q=-\Gamma H (1+r) \rho_{DE}$, equation (\ref{rhoconservde}) can be integrated to obtain 
\begin{equation}\label{lnrho3n}
 \mathrm{ln} \left(\frac{\rho_{DE}}{\rho_{0}}\right)=\int{\left[\Gamma H (1+r) -3H(1+w_{DE})\right]} dt
\end{equation}
which with equation (\ref{lnrho3n}) will give
\begin{equation}\label{lnrhodx3n}
\frac{d}{dx}\left[\mathrm{ln} \left(\frac{\rho_{DE}}{\rho_{0}}\right)\right]=\Gamma (1+r) -3(1+w_{DE})
\end{equation}
Further, equation (\ref{rhoconservm}) with $Q=-\Gamma H (1+r)\rho_{DE}$ gives
\begin{equation}
\frac{1}{\rho_m}\frac{d \rho_m}{dx} = -\left( \frac{\Gamma (1+r)}{r} + 3\right)
\end{equation}
This immediately gives 
\begin{equation}\label{lnrhoomega2r3n}
\begin{split}
\frac{d}{dx}\left[\mathrm{ln} \left(\frac{\rho_{DE}}{\rho_{0}}\right)\right]=\frac{\Omega_{DE}'}{\Omega_{DE}(1-\Omega_{DE})} + \frac{1}{(1-\Omega_{DE})} \\ \left[ 2(\Omega_{m}+\Omega_{DE}-1)-\Omega_{m}\left(\frac{\Gamma (1+r)}{r} + 3\right)\right]
\end{split}
\end{equation}
Using equations (\ref{lnrhodx3n}) and   (\ref{lnrhoomega2r3n}), one gets
\begin{equation}\label{lnrhoomega3r3n}
\begin{split}
\frac{\Omega_{DE}'}{\Omega_{DE}(1-\Omega_{DE})} + \frac{1}{(1-\Omega_{DE})}  \left[ 2(\Omega_{m}+\Omega_{DE}-1)-\Omega_{m}\left(\frac{\Gamma (1+r)}{r} + 3\right)\right]\\
=\Gamma (1+r) -3(1+w_{DE})
\end{split}
\end{equation}
Using the time derivative of $\rho_{DE}$ 
and putting this expression in equation (\ref{rhoconservde}), one obtains the expression for $w_{DE}$ for case III as
\begin{equation}\label{model3wden}
w_{DE} =- 
\left(\frac{1+\Delta}{3}\right)+\frac{\Delta-2}{3}\left(\cosh y \sqrt{\frac{3 M_p^2 \Omega_{DE}}{C}}  L^{-\frac{\Delta}{2}}\right)+\frac{\Gamma (1+r)}{3}.
\end{equation}
Using (\ref{model3wden}) in equation(\ref{lnrhoomega3r3n}), one obtains the expression for evolution of interacting Barrow holographic dark energy model (using $Q=-\Gamma H (1+r)\rho_{DE}$) in an open
universe as
\begin{equation}\label{model3den}
\begin{split}
\frac{\Omega_{DE}'}{\Omega_{DE} (1-\Omega_{DE})} = (\Delta - 2)\left(1-\cosh y \sqrt{\frac{3 M_p^2 \Omega_{DE}}{C}} L^{-\frac{\Delta}{2}}\right)\\ - \frac{1}{(1-\Omega_{DE})} \left[ 2(\Omega_{m}+\Omega_{DE}-1)-\Omega_{m}\left(\frac{\Gamma (1+r)}{r}+3\right)\right]
\end{split}
\end{equation}
In a similar way starting with (\ref{rhoconservm}) and using the expression for the source term given in (\ref{case III}), we obtain 
\begin{equation}\label{model3mn}
\begin{split}
\frac{\Omega_{m}'}{\Omega_{m}} =-\left(\frac{\Gamma (1+r)}{r}+3\right)-\Omega_{DE} (\Delta - 2)\left(1-\cosh y \sqrt{\frac{3 M_p^2 \Omega_{DE}}{C}} L^{-\frac{\Delta}{2}}\right)\\ - \left[ 2(\Omega_{m}+\Omega_{DE}-1)-\Omega_{m}\left(\frac{\Gamma (1+r)}{r}+3\right)\right]
\end{split}
\end{equation}
Differential equations (\ref{model3den}) and (\ref{model3mn}) will govern the evolution of Barrow holographic dark energy model in an open universe for the particular choice of $Q$ given in (\ref{case III}). 
\section{Cosmological behavior} 
\label{Cosmological behavior}
Observational cosmology plays a crucial role in developing precise cosmological frameworks by constraining theoretical models with observational data. In this section, we emphasize the importance of observational cosmology in developing precise cosmological frameworks by constraining various parameters of the interacting Barrow holographic dark energy (IBHDE) model, namely, $\Omega_{k0}$, $\Gamma$, $\Delta$, in closed and open universe for the three different interacting scenarios discussed earlier.
 \subsection{$\mathrm{Case ~I ~~:~~} Q=-\Gamma H \rho_{DE}$}\label{modelIda}
 Equations (\ref{model1de}) and 
(\ref{model1den}) determine the behavior of the dark-energy density parameters as a function of $x = \ln a$, for positive and negative spatial curvature cases  respectively. Similarly, equations (\ref{model1m}) and (\ref{model1mn}) determine the behavior of the 
density parameters for the matter sector for positive and negative curvature. These evolution equations can be represented conveniently in terms of the redshift parameter $z$ through  $x = \ln a =
-{\rm ln}(1 + z)$  with the present value of the  scale factor $a_0$ set to unity. Equations (\ref{model1de}), (\ref{model1m}), (\ref{model1den}) and (\ref{model1mn})  are solved numerically
imposing the initial conditions, $\Omega_m (x=-\ln(1+z)=0) \equiv \Omega_{m0}$, $\Omega_{DE} (x=-\ln(1+z)=0) \equiv \Omega_{DE0}$ and 
$\Omega_k (x=-\ln(1+z)=0) \equiv \Omega_{k0}$ in agreement with recent observations \cite{Aghanim2020}.
\begin{figure}[htbp]
\centering
\includegraphics[width=0.44\columnwidth]{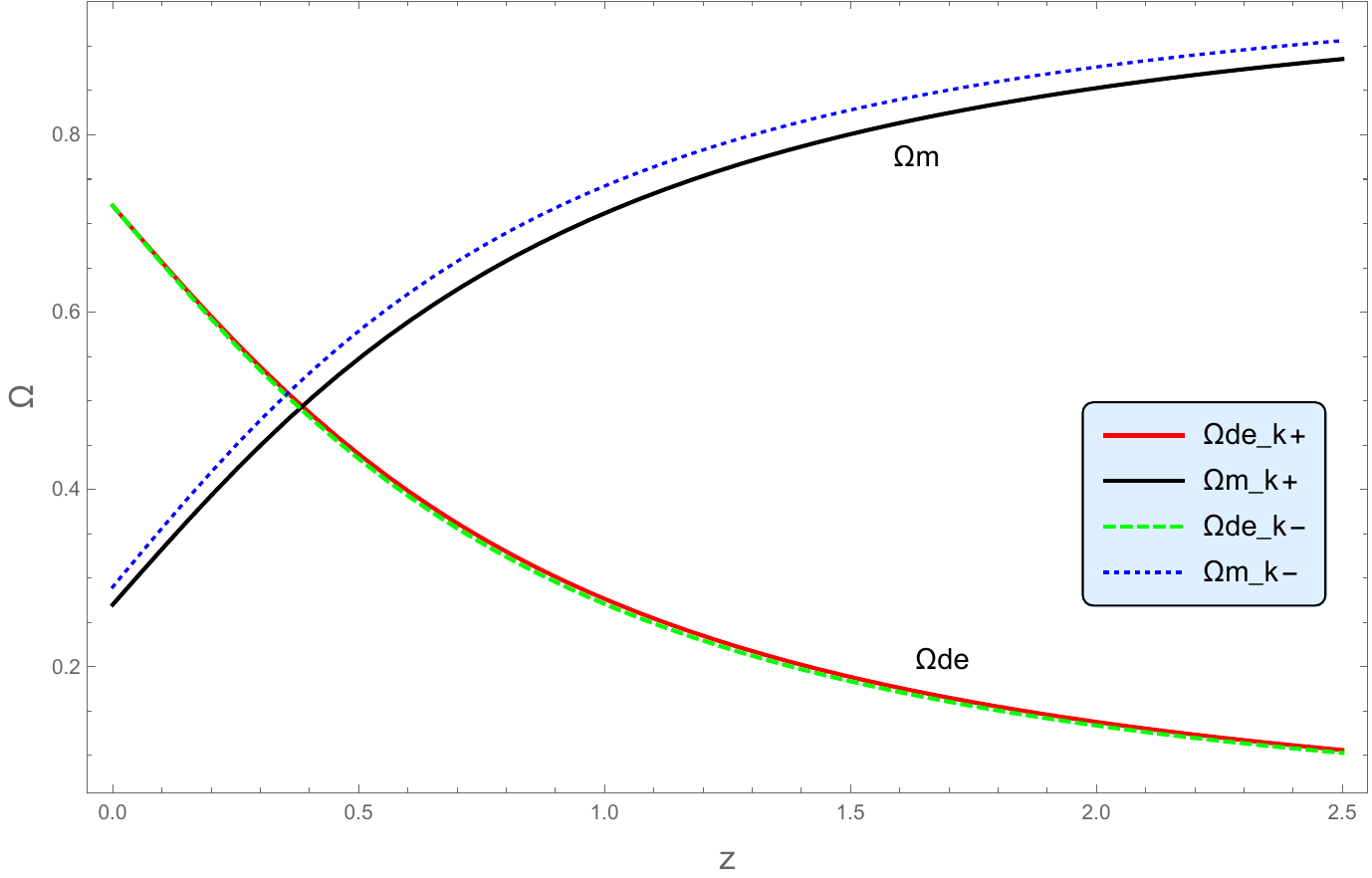}
\qquad
\includegraphics[width=0.44\columnwidth]{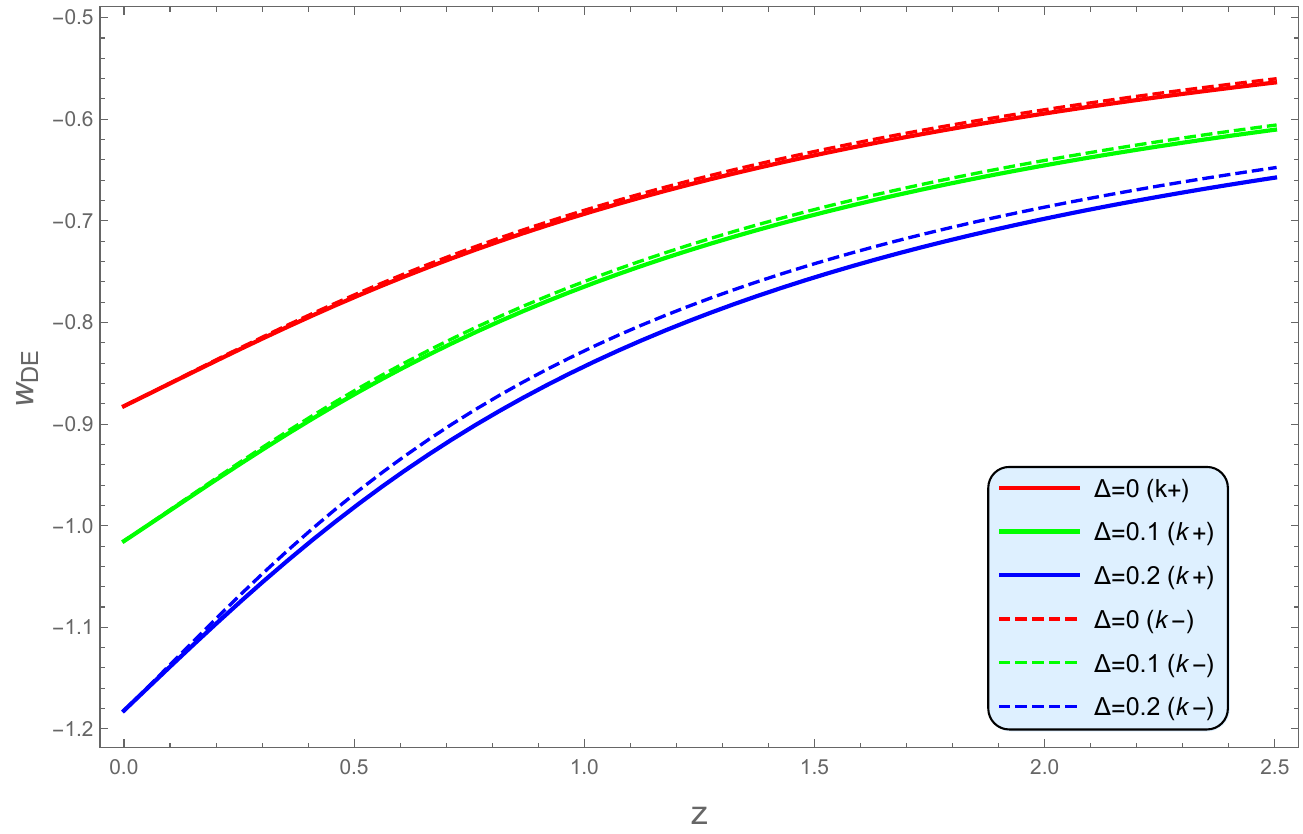}
\caption{Left panel shows the evolution of the density parameters $\Omega_m (z)$ and $\Omega_{de} (z)$ for  matter and IBHDE model with $Q=-\Gamma H \rho_{DE}$, $\Delta=0.1$, $\Gamma=0.05$ and ${C}=3$ for both closed ($k=+1$) and open universe ($k=-1$). The right panel depicts the evolution of the dark energy equation of state parameter $w_{DE}(z)$ for various  $\Delta$ values.}
\label{model1diffdelta}
\end{figure}
Left panel of figure \ref{model1diffdelta} represents the evolution of matter and dark energy density parameters $\Omega_{m}(z)$ and $\Omega_{DE}(z)$ in the case of closed universe (denoted by $\Omega_m k+$ and $\Omega_{de} k+$) and open universe cases (denoted by $\Omega_m k-$ and $\Omega_{de} k-$) for specific values of the Barrow exponent $\Delta$, the coupling strength $\Gamma$ and other model parameters. As evident from the plot, the interacting model depicts the usual thermal history exhibiting the transition from matter to dark energy dominance happening at around $z \sim 0.4$ for positive curvature and around $z \sim 0.35$ for negative curvature case. For the density parameter plots, we have chosen $\Delta=0.1$, $\Gamma=0.05$, ${C}=3$, $\Omega_{DE0}\approx 0.72$, $\Omega_{m0}\approx 0.29$ for both closed ($k=+1$) and open universe ($k=-1$) models. Further $\Omega_{k0}= 0.01$ for positive curvature universe ($k=+1$) and $\Omega_{k0}=-0.01$ for negative curvature universe ($k=-1$) has been considered. In order to gain a more detailed and clear insight about the cosmological behaviour of the proposed interacting Barrow holographic dark energy model, we plot $w_{DE}(z)$ for both $k=+1$ and $k=-1$ cases for different values of $\Delta$ which has been shown in the right panel of figure \ref{model1diffdelta}. The symbols $k+$ and $k-$ represent the positive and negative curvature models respectively. 
As one can see, for $\Delta =0$, $w_{DE}(z)$ lies completely in the quintessence regime. However, with the increase in the value of $\Delta$, $w_{DE}(z)$ tend to approach the phantom regime for both closed as well as open universe models. 
To be more specific, $w_{DE}(z)$ just crosses the phantom-divide line at $z \sim 0.5$ for $\Delta=0.1$ whereas for $\Delta=0.2$, it enters the phantom regime much earlier (at around $z \sim 0.5$). A similar behaviour was obtained for the non-interacting Barrow holographic dark energy model \cite{Adhikary:2021} but the phantom-divide crossing was realized much earlier. So the interaction between dark energy and dark matter provides a phantom-divide crossing for comparative higher values of $\Delta$.
\begin{figure}[htbp]
\centering
\includegraphics[width=.55\columnwidth]{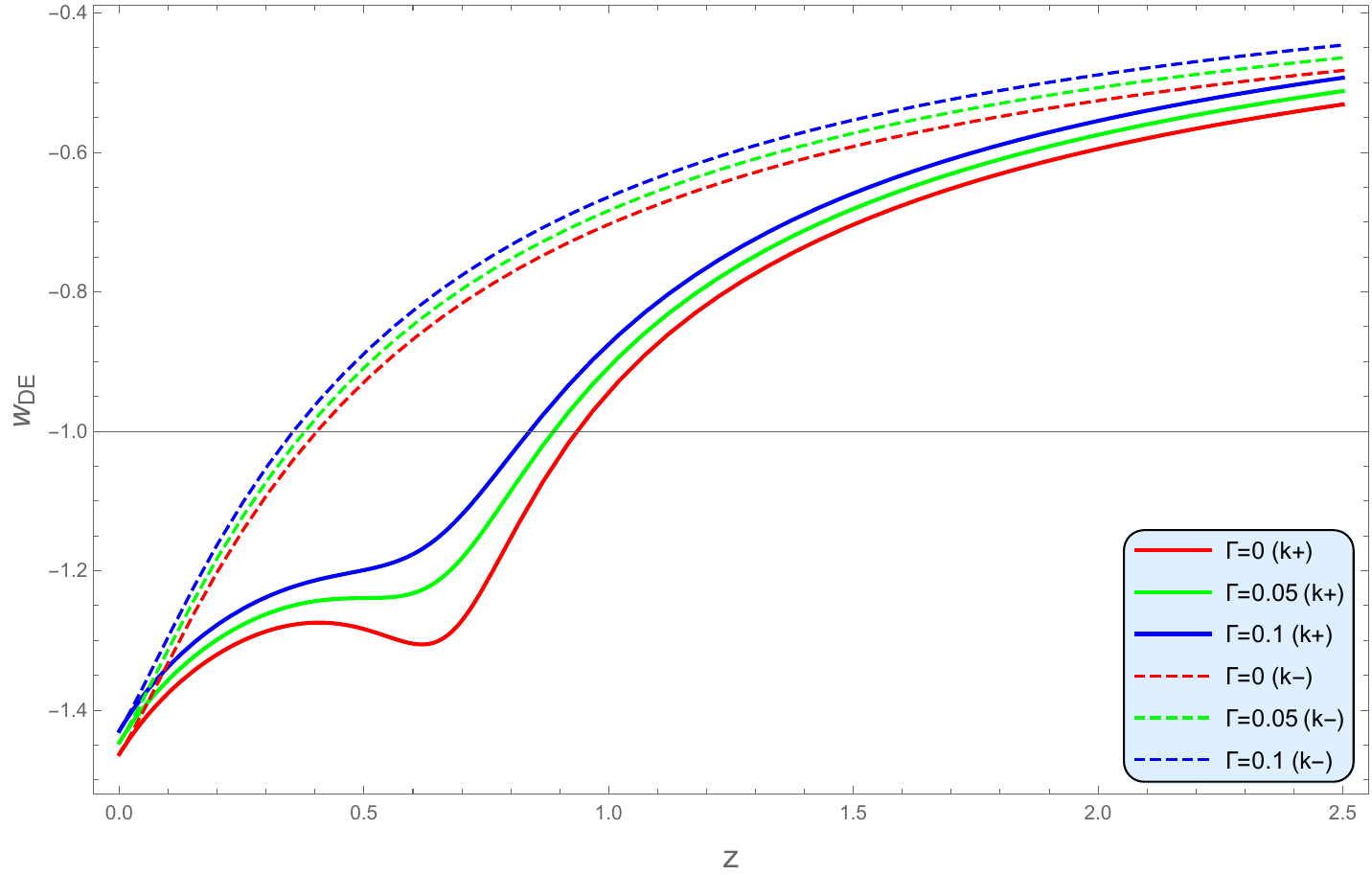}
\caption{The evolution of the dark energy equation of state parameter $w_{DE}(z)$ for $Q=-\Gamma H \rho_{DE}$ for different $\Gamma$ values. We have put $\Delta$ = $0.2$, $\Omega_{DE0}\approx 0.72$, $\Omega_{m0}\approx 0.29$ and $\Omega_{k0}= +0.01$ or $-0.01$ for positive and negative curvature cases respectively.} 
\label{model1diffgamma}
\end{figure}
We have also plotted $w_{DE}(z)$ keeping $\Delta$ fixed ($\Delta =0.2$) and considering different strengths of interaction $\Gamma$ for open  and closed universe models as shown in figure \ref{model1diffgamma}. This will help us to understand, how the evolution of the universe is affected by the strength of the interaction or combinations of these model parameters. It has been found that for both open and closed cases, the evolutionary behaviour is phantom at present. However, the closed universe models are found to enter the phantom regime earlier as compared to open models. In particular, for $\Gamma=0$ the dark energy EOS parameter for $k=+1$ case crosses the phantom divide line at $z \sim 0.9$ whereas for $k=-1$  the phantom crossing happens at $z \sim 0.4$ with a higher Barrow exponent value ($\Delta=0.2$). For non-interacting case, the phantom-divide crossing was realized for $\Delta > 0.03$ \cite{Adhikary:2021}, but for this particular interacting model, the phantom-divide crossing has been obtained for comparatively higher values of $\Delta$. As evident from figure \ref{model1diffdelta}, with zero coupling strength ($\Gamma = 0$) and zero Barrow exponent ($\Delta = 0$), the Universe lies totally in the quintessence regime; but with small nonzero values of $\Delta$, the universe is found to cross phantom divide crossing for both positive and negative curvature cases and becomes more and more phantom with increasing values of $\Gamma$ or $\Delta$. These results are consistent with the standard non-interacting holographic dark energy models \cite{Saridakis:2020PRD, Adhikary:2021}.
%%%%%%%%%%%%%%%%%%%%%%%%%%%%%%%%%%%%%%%%%%%%%%%%%%%%
\subsection{$\mathrm{Case ~II ~~:~~} Q=-\Gamma H r\rho_{DE}$}\label{modelIIda}
\begin{figure}[htbp]
\begin{center}
\includegraphics[width=0.49\columnwidth]{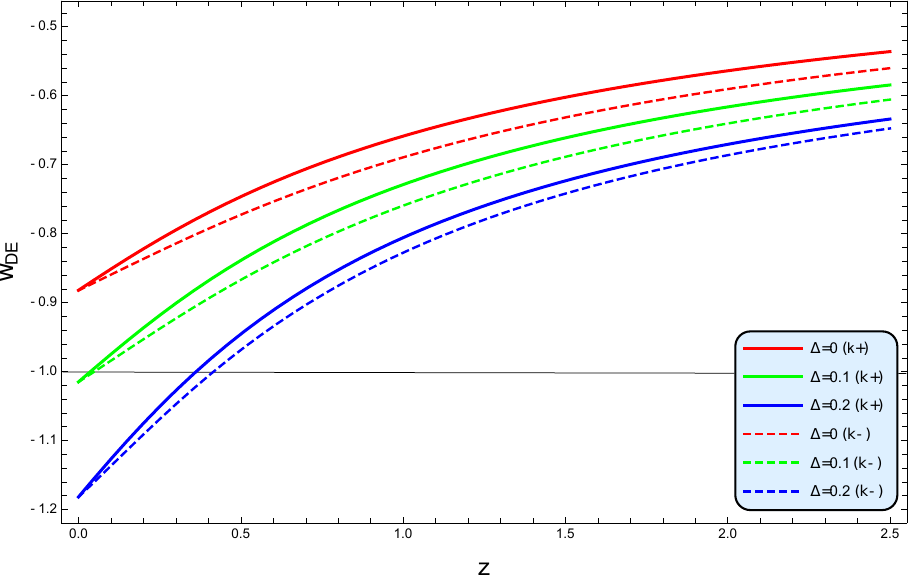}
\includegraphics[width=0.49\columnwidth]{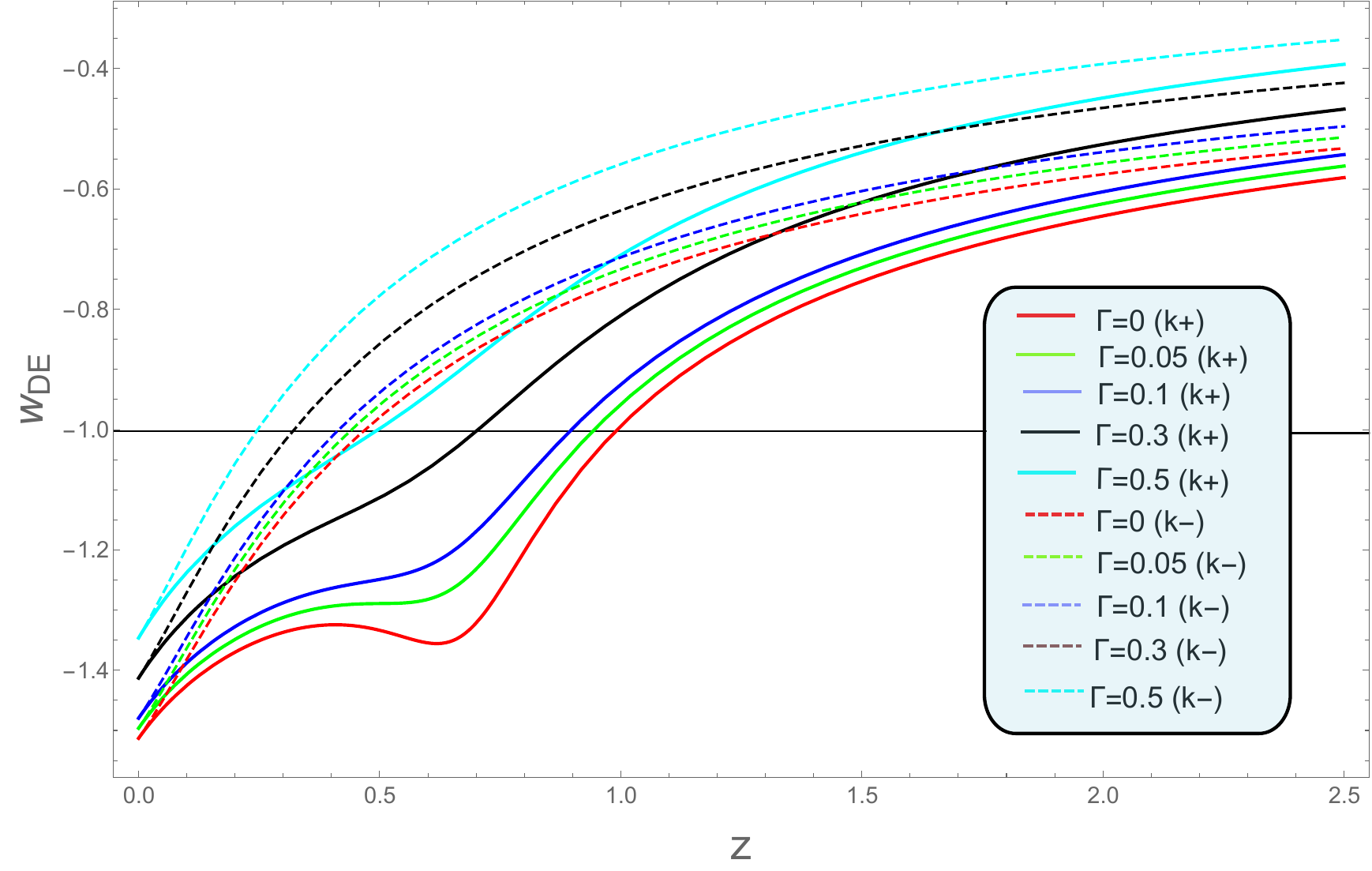}
\caption{\em
Evolution of $w_{DE}(z)$ for model II ($Q=\Gamma Hr \rho_{DE}$) for closed ($k=+1$) and open universe ($k=-1$) cases for various  $\Delta$ values and $\Gamma$.
We have imposed $\Omega_{DE0}\approx 0.72$, $\Omega_{m0}\approx 0.27$, $\Omega_{k0}= 0.01$ for positive curvature and $\Omega_{k0}= -0.01$ for negative curvature universe.  
} 
\label{model2diffdelta}
\end{center}
\end{figure}
For this particular form of interaction also, we have plotted the equation of state parameter $w_{DE}(z)$ for both $k = +1$ and $k = -1$ cases for different values of $\Delta$ (left panel of figure \ref{model2diffdelta}) as well as for different values of $\Gamma$ (right panel of figure \ref{model2diffdelta}). For the left panel we have set $\Omega_{DE0}\approx 0.72$, $\Omega_{m0}\approx 0.27$ and $\Gamma =0.05$ whereas for the right panel we have considered $\Omega_{DE0}\approx 0.72$, $\Omega_{m0}\approx 0.27$ and $\Delta =0.2$. The evolution of the density parameters ($\Omega_m$ and $\Omega_{DE}$) for  matter and interacting Barrow holographic dark energy for model II ($Q=\Gamma Hr \rho_{DE}$) exhibits a similar behaviour as model I. It has been observed that for very small values of the strength of interaction $\Gamma$, the equation of state parameter $w_{DE}(z)$ exhibits a very similar behaviour as case I for small $\Delta$ values. However, higher values of the interaction strength $\Gamma$, the phantom-divide crossing occurs at later times. As before, the plot of $w_{DE}(z)$ for different strengths of interaction $\Gamma$ for open  and closed universe models (right panel of figure \ref{model2diffdelta}) shows that the evolution becomes more and more phantom with increasing values of $\Gamma$ or $\Delta$.

\subsection{$\mathrm{Case ~III ~~:~~} Q=-\Gamma H (1+r)\rho_{DE}$}\label{modelIIIda}
\begin{figure}[ht]
\begin{center}
\includegraphics[width=0.49\columnwidth]{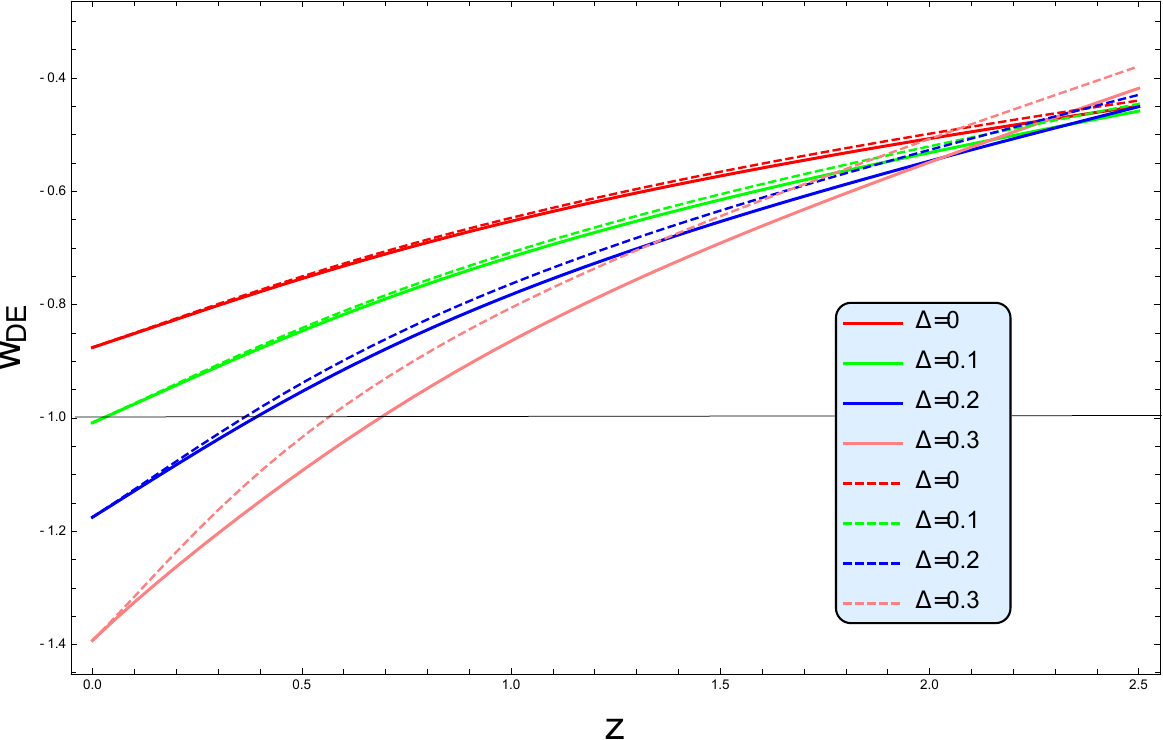}
\includegraphics[width=0.49\columnwidth]{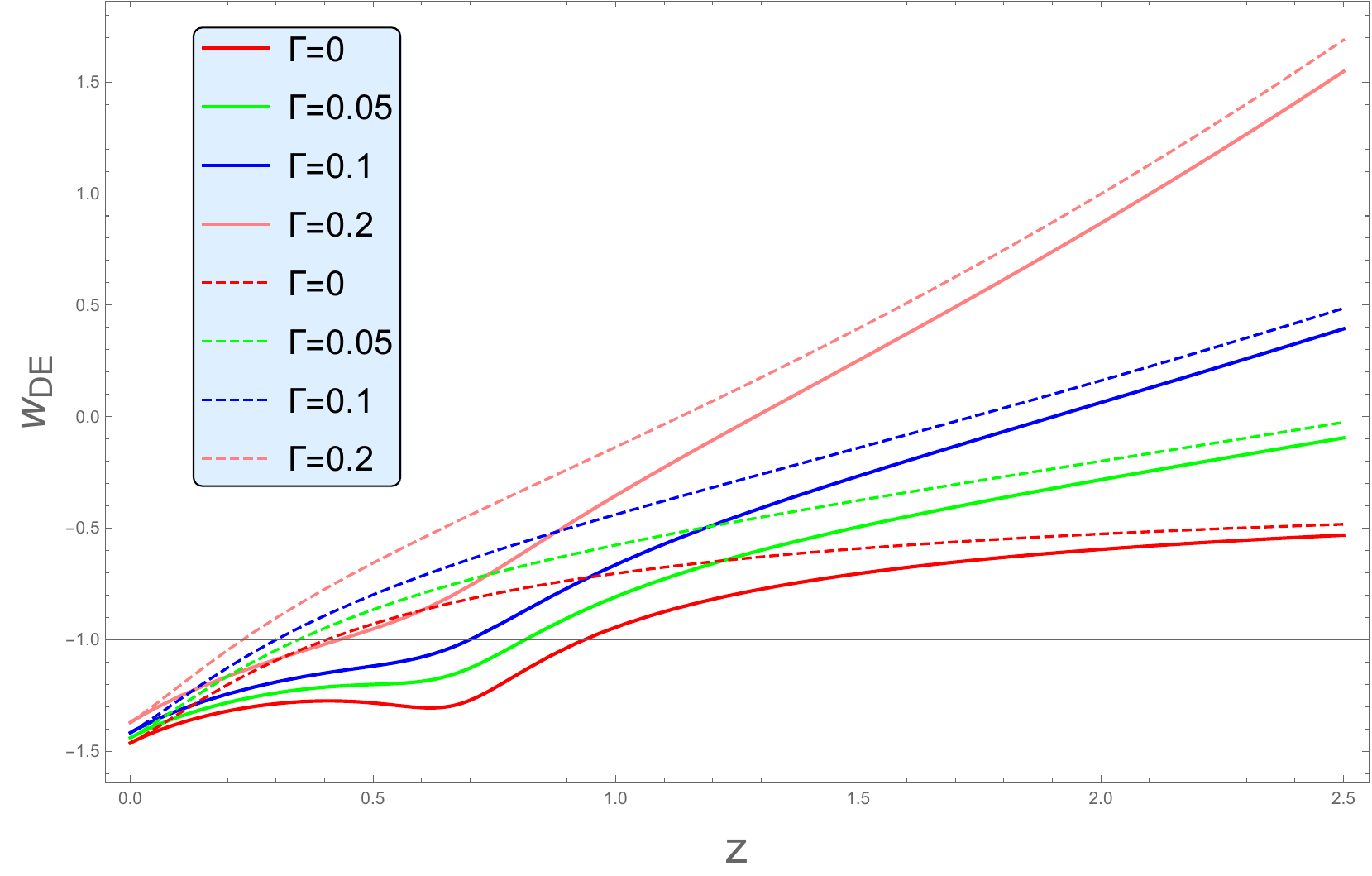}
\caption{\em
Evolution of $w_{DE}(z)$ for model III with $Q=-\Gamma H(1+r) \rho_{DE}$ for closed ($k=+1$) and open universe ($k=-1$) cases for different values of $\Delta$ and $\Gamma$. We have imposed $\Omega_{DE0}\approx 0.72$, $\Omega_{m0}\approx 0.27$, $\Omega_{k0}= 0.01$ for positive curvature and $\Omega_{k0}= -0.01$ for negative curvature universe.} 
\label{model3diffdelta}
\end{center}
\end{figure}
For this particular choice, the interaction term is considered to be dependent on the total energy density of the universe comprising of both dark energy and dark matter sectors as given in equation (\ref{case III}). Like previous cases, for this form of interaction also, we have plotted the equation of state parameter $w_{DE}(z)$ for both $k = +1$ and $k = -1$ cases for different values of $\Delta$ (left panel of figure \ref{model3diffdelta}) as well as for different values of $\Gamma$ (right panel of figure \ref{model3diffdelta}). For both the plots, we have set $\Omega_{DE0}\approx 0.72$, $\Omega_{m0}\approx 0.27$. For the left panel we have set $\Gamma =0.05$ and varied $\Delta$ whereas for the right panel we have fixed $\Delta =0.2$ and varied $\Gamma$. 

%%%%%%%%%%%%%%%%%%%%%%%%%%%%%%%%%%%%%

\if else
\begin{figure}[ht]
\begin{center}
\includegraphics[width=0.48\columnwidth]{different_delta_H(1+r)_rhode_omega.png}
\includegraphics[width=0.48\columnwidth]{different_delta_H(1+r)_rhode_wde.png}
\caption{\em
Upper graph: The evolution of the density parameters for  matter and    Interacting Barrow holographic dark energy for model III ($Q=\Gamma H(1+r) \rho_{DE}$), as a function of the redshift $z$, in the case of both closed universe ($k=+1$) and open universe ($k=-1$), for $\Delta=0.1$, $\Gamma=0.05$ and ${C}=3$, in    $M_p^2=1$ units. Lower graph: The evolution of the   dark-energy equation-of-state parameter $w_{DE}(z)$ for model III ($Q=\Gamma H(1+r) \rho_{DE}$), as a function of the redshift $z$, in the case of both closed universe ($k=+1$) and open universe ($k=-1$), for various  $\Delta$ values.We have imposed $\Omega_{DE0}\approx0.72$, $\Omega_{m0}\approx0.27$  and $\Omega_{k0}= 0.01$ for positive curvature universe ($k=+1$) and$\Omega_{DE0}\approx0.72$, $\Omega_{m0}\approx0.29$  and $\Omega_{k0}=-0.01$ for negative curvature universe ($k=-1$) at present. } 
\label{model3diffdelta}
\end{center}
\end{figure}

\begin{figure}[ht]
\begin{center}
\includegraphics[width=0.48\columnwidth]{different_gamma_H(1+r)_rhode_omega.png}
\includegraphics[width=0.5\columnwidth]{different_gamma_H(1+r)_rhode_wde.png}
\caption{\em Upper graph: The evolution of the density parameters for  matter and    Interacting Barrow holographic dark energy for model III ($Q=\Gamma H(1+r) \rho_{DE}$), as a function of the redshift $z$, in the case of both closed universe ($k=+1$) and open universe ($k=-1$), for $\Delta=0.4$, $\Gamma=0.1$ and ${C}=3$, in    $M_p^2=1$ units. Lower graph: The evolution of the   dark-energy equation-of-state parameter $w_{DE}(z)$ for model III ($Q=\Gamma H(1+r) \rho_{DE}$), as a function of the redshift $z$, in the case of both closed universe ($k=+1$) and open universe ($k=-1$), for various  $\Gamma$ values. We have imposed $\Omega_{DE0}\approx0.72$, $\Omega_{m0}\approx0.27$  and $\Omega_{k0}= 0.01$ for positive curvature universe ($k=+1$) and$\Omega_{DE0}\approx0.72$, $\Omega_{m0}\approx0.29$  and $\Omega_{k0}=-0.01$ for negative curvature universe ($k=-1$) at present. } 
\label{model3diffgamma}
\end{center}
\end{figure}

\begin{figure}[ht]
\begin{center}
\includegraphics[width=0.84\columnwidth]{wDE_vs_z_k_0.01_negative.pdf}
\caption{\em
The evolution of the dark-energy equation-of-state 
parameter for   Barrow 
holographic dark energy, as a function of the redshift 
$z$, in the case of an open universe ($k=-1$), for  various  $\Delta$ 
values.
We have imposed 
 $\Omega_{DE0}\approx0.72$, $\Omega_{m0}\approx0.29$  and $\Omega_{k0}= -
0.01$ at present. The vertical line marks the present time $z=0$.  }
\label{figwDEnegativek}
\end{center}
\end{figure}
\else
 
%%%%%%%%%%%%%%%%%%%%%%%%%%%%%%%%%%%%%
The evolution of $\Omega_{m}(z)$ and $\Omega_{DE}(z)$ are found to be similar to the previous cases exhibiting the usual thermal history of the universe. In figure \ref{model3diffdelta}, we can see that for the positive curvature case, with the increase in the value of $\Delta$, $w_{DE}$ becomes more and more negative. For smaller values of $\Delta$, the evolution is in quintessence region but for higher values of $\Delta$, it becomes phantom. On the other hand, for negative curvature case, for all the $\Gamma$ values, the universe is currently in the phantom regime. Hence one can conclude that for the case of open spatial geometry, the phantom regime is more favorable, contrary to the case of spatially flat universe \cite{Saridakis:2020zol} or the positive curvature case analyzed 
above. 

\section{Observational Constraints}
\label{Observational Constraints}
In this section we try to obtain observational constraints on the model parameters $\Omega_{k0}$, $\Gamma$ and $\Delta$ by fixing the values of  $\Omega_{m0}$ and $\Omega_{DE0}$. For this analysis, first we fix the value of $\Delta$ at $\Delta=0.1$ and try to obtain constraints on the model parameters $\Omega_{k0}$ and $\Gamma$ corresponding to different datasets for both positive and negative curvature universe. This is justified as we have seen from the previous figures that the equation of state parameters exhibit more or less similar profiles for different values of $\Delta$ for the interacting Barrow holographic dark energy model. We also try to obtain constraints on the Barrow exponent $\Delta$ from observational data and for that we carry out the analysis in the $\Gamma$ - $\Delta$ parameter.
%as well as in the $\Omega_{k0}$ - $\Delta$ parameter space.}} 
To constrain the model parameters using the latest cosmological data, we employ
Markov Chain Monte Carlo (MCMC) analysis \cite{LewisMCMC} using the following publicly available observational datasets:  
\begin{itemize}
    \item {\bf{Cosmic Chronometer data:}} In Cosmic Chronometry, the Hubble parameter $H(z)$ at different redshifts is usually determined through two approaches: (i) by extracting $H(z)$ from line-of sight of BAO data \cite{Chuang_2016, Bautista_2017} and (ii) by estimating $H(z)$ via the method of differential age (DA) of galaxies \cite{Simon:2005, Jimenez_2002, Ratsimbazafy_2017, zhang:2014} which relies on the relation
    $$H(z) = -\frac{1}{1+z} \frac{dz}{dt}$$
    where $\frac{dz}{dt}$ is approximated by determining the time interval $\Delta t$ corresponding to a given $\Delta z$ \cite{Ratsimbazafy_2017}. \\
In this work we have used the 57 data points of Hubble parameter measurements in the
 redshift range $0.07 \le z \le 2.36$ of which 31 points have been measured via the method of differential age (DA) and the remaining 26 through BAO and other methods \cite{mhamdi2024}. The $\chi^2$ function for the cosmic chronometer dataset is defined as
\begin{equation}
\chi^2_{CC} = \sum^{57}_{i=1}\frac{[{h}^{obs}(z_{i}) - 
{h}^{th}(z_{i})]^2}{\sigma^2_{H}(z_{i})} \end{equation}
where ${h} = \frac{H(z)}{H_{0}}$ is the normalized Hubble parameter. The superscripts {\it obs} and {\it th} refer to the observational values and the corresponding theoretical values respectively and $\sigma_{H}$ indicates the error for each data point. 
 
\item {\bf{Pantheon data:}} We have also used the Pantheon compilation data which is a collection of SNIa data in the redshift range $0.01 < z < 2.26$ with 1048 data points \cite{scolnic2018complete,peng2023pantheon}. The $\chi^2$ function for Pantheon dataset is given by \citep{scolnic2018complete}
\begin{equation}
 \chi^{2}_{Pantheon}=\sum_{i,j=1}^{1048}(\mu^{th}-\mu^{obs})_{i} (C_{Pantheon})^{-1}_{ij} (\mu^{th}-\mu^{obs})_{j}\\ 
\end{equation}
where  $\mu^{obs}$, $\mu^{th}$ represent the observed distance modulus and the corresponding theoretical value respectively and $(C_{Pantheon})^{-1}$ corresponds to the inverse of covariance matrix for Pantheon sample. The theoretical distance modulus is given by
\begin{equation}
\mu_{th}(z)=5log_{10}\frac{d_{L}(z)}{1 Mpc}+25,
\end{equation}
where $d_{L}(z)$ represents the luminosity distance which can be obtained by integrating the expression
 \begin{equation}
d_{L}(z,\theta_{s}) = (1+z)\int_{0}^{z}\frac{dz'}{H(z',\theta_{s})},
 \end{equation}
where $H(z)$ is the expression of the Hubble parameter for the cosmological model and $\theta_{s}$ is the parameter space of the cosmological model.\\
\if else
The $\chi^2$ for supernova data is defined as \cite{NesserisPRD:2005}
\begin{equation}
\chi^2_{SN}= A - \frac{B^2}{C},
\end{equation}
with $A$, $B$ and $C$   defined as  
\begin{eqnarray}
A = \sum^{580}_{i=1} \frac{[{\mu}^{obs}(z_{i}) - 
{\mu}^{th}(z_{i})]^2}{\sigma^2_{i}},
\end{eqnarray}
\begin{eqnarray}
B = \sum^{580}_{i=1} \frac{[{\mu}^{obs}(z_i) - 
{\mu}^{th}(z_{i})]}{\sigma^2_{i}},
\end{eqnarray}
and
\begin{equation}
C = \sum^{580}_{i=1} \frac{1}{\sigma^2_{i}}
\end{equation}
where $\mu^{obs}$ represents the observed distance modulus at a particular 
redshift,  $\mu^{th}$ the corresponding theoretical value and $\sigma_{i}$ 
represents the uncertainty in the distance modulus.\\
\else
The total $\chi^2$ for these combined datasets
is given by
$$\chi^2_{total} =\chi^2_{CC} + \chi^2_{Pantheon} $$
\end{itemize}
%%%%%%%%%%%%%%%%%%%%%%%%%%%%%%
\begin{figure}[!h]
\begin{center}
\includegraphics[width=0.45\columnwidth]
{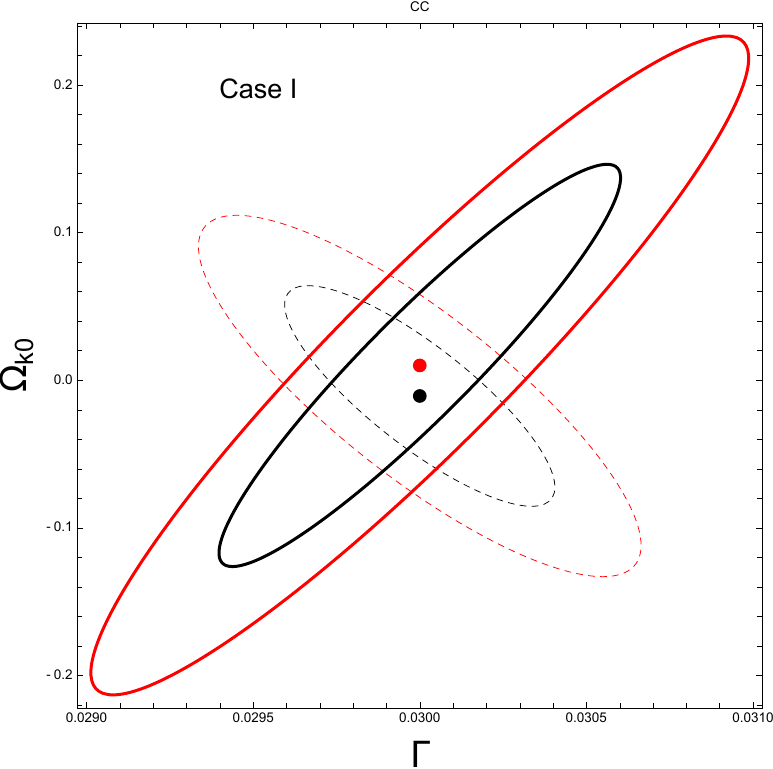} 
\includegraphics[width=0.45\columnwidth]
{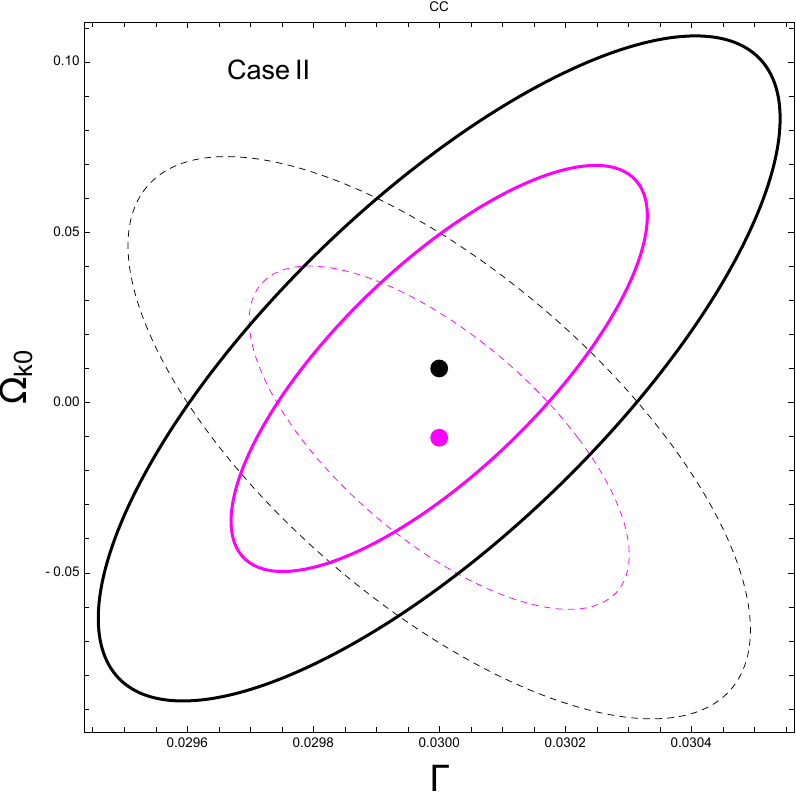}
\includegraphics[width=0.45\columnwidth]
{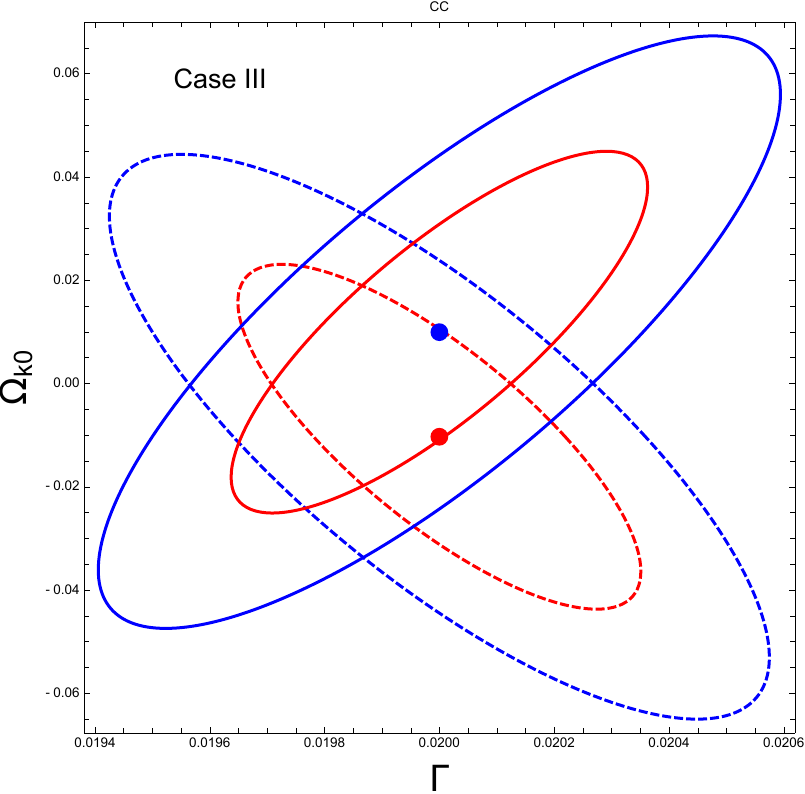}
\caption{\em 
The $1\sigma$ and $2\sigma$ likelihood contours in $\Omega_{k0} - \Gamma$ parameter space for all three cases corresponding to Cosmic Chronometer dataset. For all the plots, the solid lines indicate the contours for positive curvature and the dashed contours represent the same for negative curvature. The dots represent the corresponding best-fit values.}
\label{CC_H_rhode_fig}
\end{center}
\end{figure}
%%%%%%%%%%%%%%%%%%%%%%%%%%%%%%%%%
We present the $1\sigma$ and $2\sigma$ 
confidence contours in the $\Omega_{k0} -\Gamma$ parameter space for $CC$ dataset corresponding to the three different interacting forms considered in equations (\ref{case I}) - (\ref{case III}). Figure \ref{CC_H_rhode_fig} depicts the scenario for the three interacting cases referred to as Case I, Case II and Case III respectively. Throughout the text, for each panel, the solid lines will indicate the confidence contours for positive curvature and the dashed contours will represent the same for negative curvature. The dots will represent the corresponding best-fit values. In figure \ref{CC_H_rhode_fig}, for the first panel (Case-I), the black and red solid lines correspond to the $1\sigma$ and $2\sigma$ confidence contours respectively for the positive curvature case and the black and red dashed lines correspond to the same for the negative curvature case. The red dot represents the best-fit value for positive curvature and the black dot represents the best-fit value for negative curvature. In the second panel of figure \ref{CC_H_rhode_fig}, similarly, the magenta and black solid and dashed lines correspond to the $1\sigma$ and $2\sigma$ confidence contours respectively for the positive and negative curvature.  The corresponding best-fit values are indicated by the black and magenta dots. The third panel corresponds to case III where the confidence contours are represented by red and blue solid and dashed lines respectively for CC dataset.
%%%%%%%%%%%%%%%%%%%%%%%%%%%%%%%%%%%%%
\begin{figure}[!h]
\begin{center}
\includegraphics[width=0.45\columnwidth]
{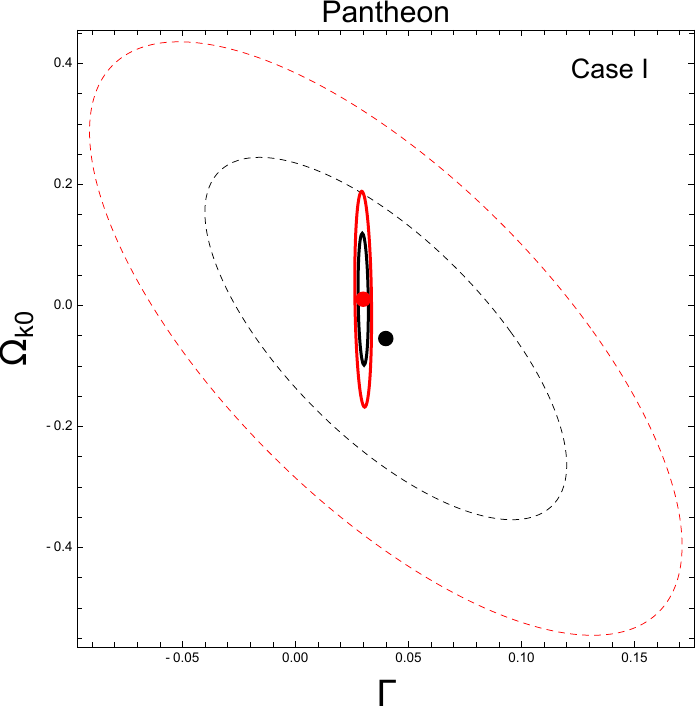} 
\includegraphics[width=0.45\columnwidth]
{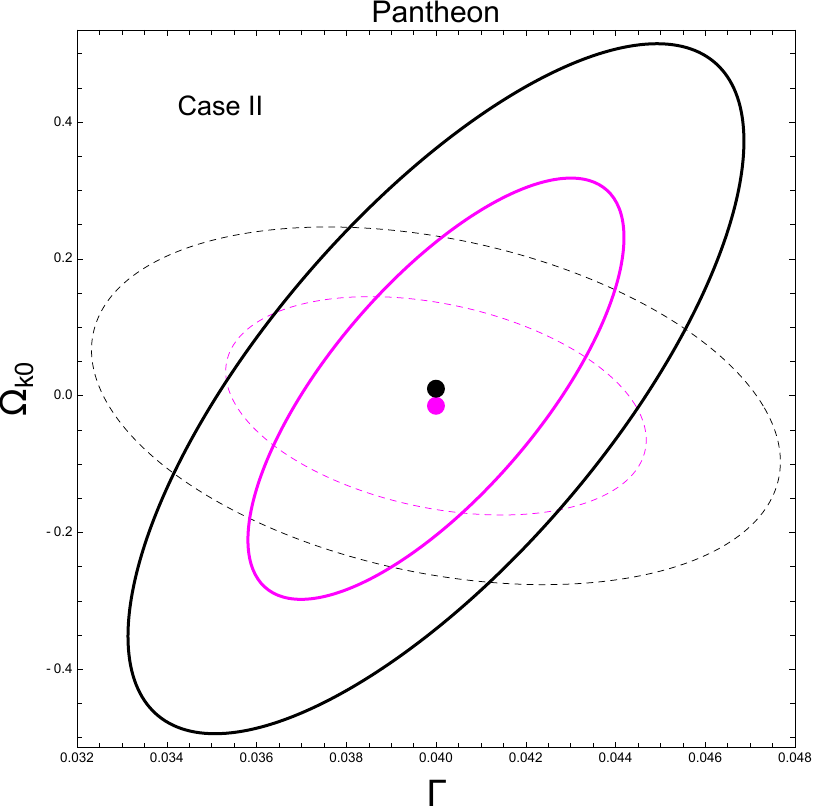}
\includegraphics[width=0.45\columnwidth]
{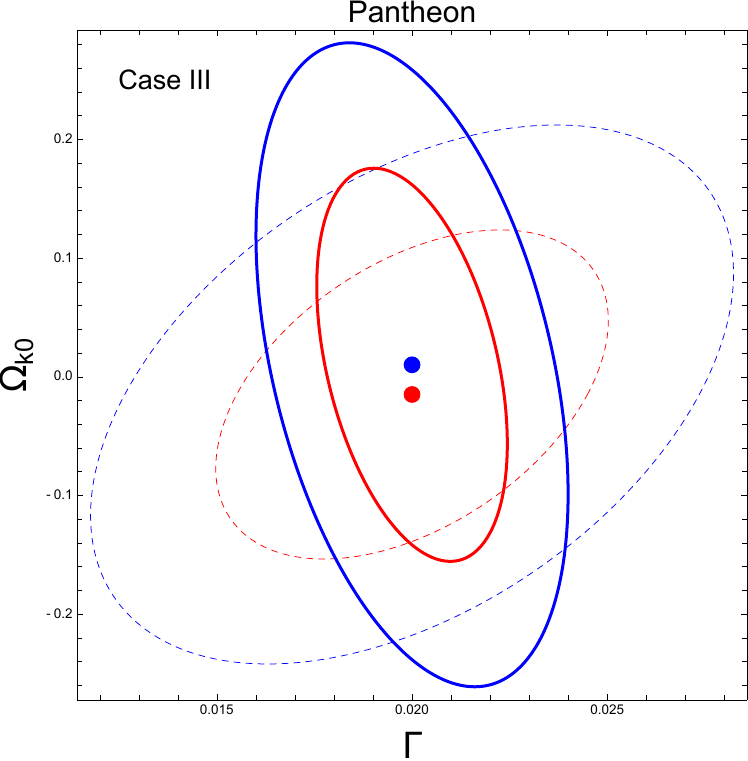}
\caption{\em 
The $1\sigma$ and $2\sigma$ confidence contours for all three cases in $\Omega_{k0} - \Gamma$ parameter space corresponding to Pantheon dataset. As mentioned, the solid lines indicate the contours for positive curvature and the dashed lines represent contours for negative curvature.}
\label{SN_H_rhode_fig}
\end{center}
\end{figure}
%%%%%%%%%%%%%%%%%%%%%%%%%%%%%%%%%%%%%
Clearly we can see that for all the three cases, the cosmic chronometer data puts tighter constraints on the parameters $\Gamma$ and $\Omega_{k_0}$ for negative curvature case. Also for case III, the constraints are tighter as compared to case I or case II. The strength of interaction as well as the curvature contribution come out to be nonzero which indicates that a non-flat interacting scenario is preferred by observational data. It is also evident that the best-fit values of the concerned parameters $\Gamma$ and $\Omega_{k_0}$ happen to be very small even at the $1\sigma$ and $2\sigma$ confidence levels. Thus comparison of these three models indicate that for the interaction term, which is phenomenologically chosen as the combination of dark matter and dark energy components, the parameter $\Omega_{k_0}$ happens to be more constrained with justified values. Also, as evident from the plots, for all the three cases, $\Gamma$ and $\Omega_{k_0}$ are positively correlated for positive curvature and negatively correlated the negative curvature case.\\
\begin{table}
\begin{center}
\begin{tabular}{ |c|c|c|c|c|c| } 
\hline
\hline
Model & Dataset & $k$ & $\Gamma$ & $\Omega_{k0}$ & $\chi^2_{\text{min}}$ \\
\hline
\multirow{6}{4em}{Case I} 
&$CC$ & positive & $0.03$ & $0.00982$ & $67.91$ \\ 
&& negative & $0.03$ & $-0.01062$ & $57.71$ \\
\cline{2-6}
&$Pantheon$ & positive & $0.03$ & $0.01$ & $66.27$ \\ 
&& negative & $0.04$ & $-0.055$ & $38.32$ \\
\cline{2-6}
&$CC+Pantheon$ & positive & $0.03$ & $0.00982$ & $86.78$ \\ 
&& negative & $0.03$ & $-0.01062$ & $80.66$ \\
\hline
\multirow{6}{4em}{Case II} 
&$CC$ & positive & $0.03$ & $0.00984$ & $53.98$ \\ 
&& negative & $0.03$ & $-0.01026$ & $49.61$ \\
\cline{2-6}
&$Pantheon$ & positive & $0.04$ & $0.01$ & $44.42$ \\ 
&& negative & $0.04$ & $-0.015$ & $47.93$ \\
\cline{2-6}
&$CC+Pantheon$ & positive & $0.03$ & $0.00983$ & $88.85$ \\ 
&& negative & $0.03$ & $-0.01026$ & $85.20$ \\
\hline
\multirow{6}{4em}{Case III} 
&$CC$ & positive & $0.02$ & $0.00979$ & $54.16$ \\ 
&& negative & $0.02$ & $-0.01021$ & $49.64$ \\
\cline{2-6}
&$Pantheon$ & positive & $0.02$ & $0.01$ & $66.57$ \\ 
&& negative & $0.02$ & $-0.015$ & $44.42$ \\
\cline{2-6}
&$CC+Pantheon$ & positive & $0.02$ & $0.00979$ & $89.12$ \\ 
&& negative & $0.02$ & $-0.01021$ & $84.96$ \\
\hline
\hline
\end{tabular}
\end{center}
\caption{Best-fit values of $\Gamma$ and $\Omega_{k0}$ for interacting Barrow holographic dark energy in the case of non-flat universe for various datasets} \label{table1}
\end{table}
In table \ref{table1} the resulting best-fit values for the parameters $\Gamma$ and $\Omega_{k0}$ 
for the different datasets has been listed. As evident from the table, for all the datasets, the best-fit values for the strength of interaction $\Gamma$ as well as the curvature density parameter $\Omega_{k0}$ comes out to be very small, but nonzero. Also the values obtained are more or less consistent for all the datasets.

Figure \ref{SN_H_rhode_fig} and figure \ref{CCSN_H_rhode_fig} represent the $1\sigma$ and $2\sigma$ confidence contours for Pantheon data and CC + Pantheon 
datasets respectively where case I, case II and case III refers to the three different interacting forms considered in equations
(\ref{case I}) - (\ref{case III}). The same colour coding and contour profiles as figure {\ref{CC_H_rhode_fig}} have been used for both these plots. For all the plots shown in figure \ref{CC_H_rhode_fig}, figure \ref{SN_H_rhode_fig} and figure \ref{CCSN_H_rhode_fig}, the analysis has been carried out considering $\Delta = 0.1$.  

%%%%%%%%%%%%%%%%%%%%%%%%%%%%%%%%%%%%%
\begin{figure}[!h]
\begin{center}
\includegraphics[width=0.45\columnwidth]
{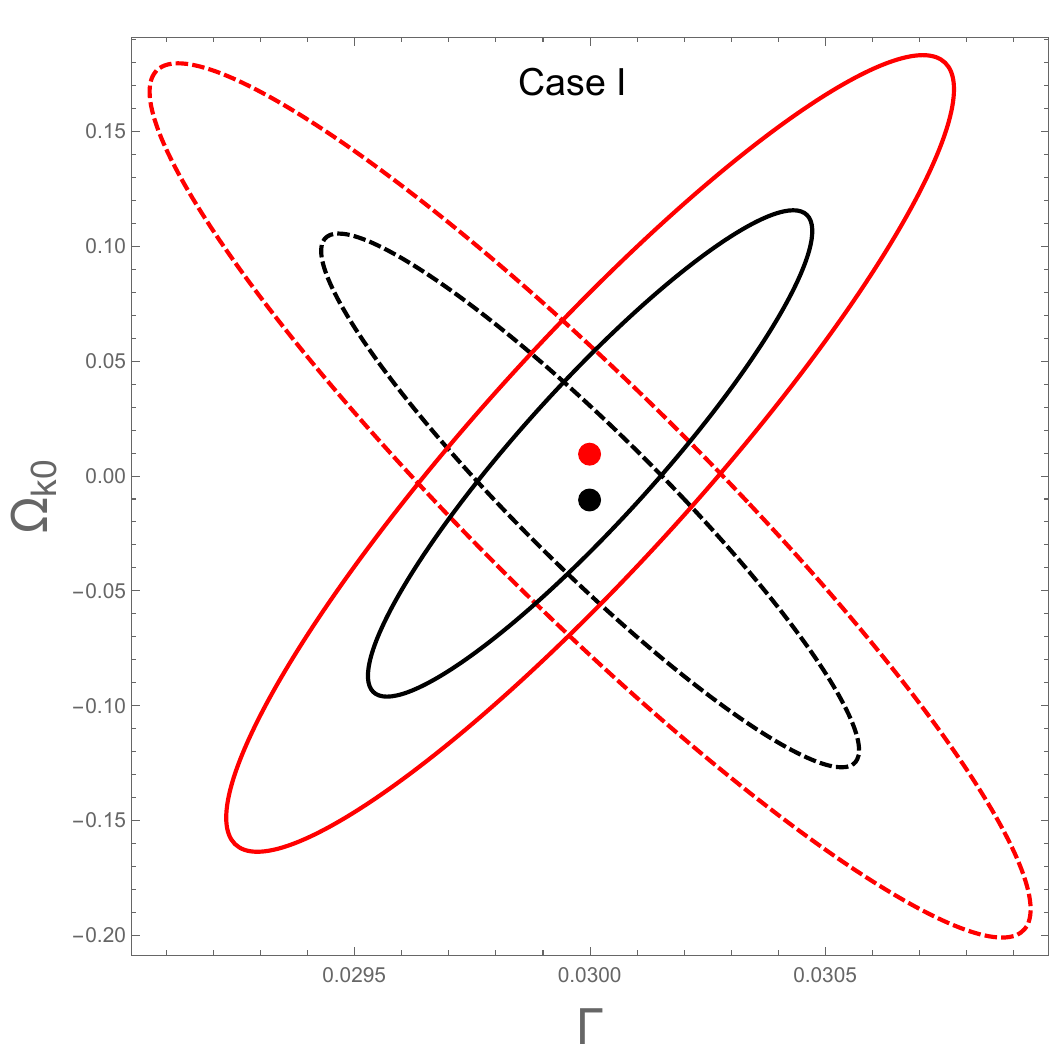} 
\includegraphics[width=0.45\columnwidth]
{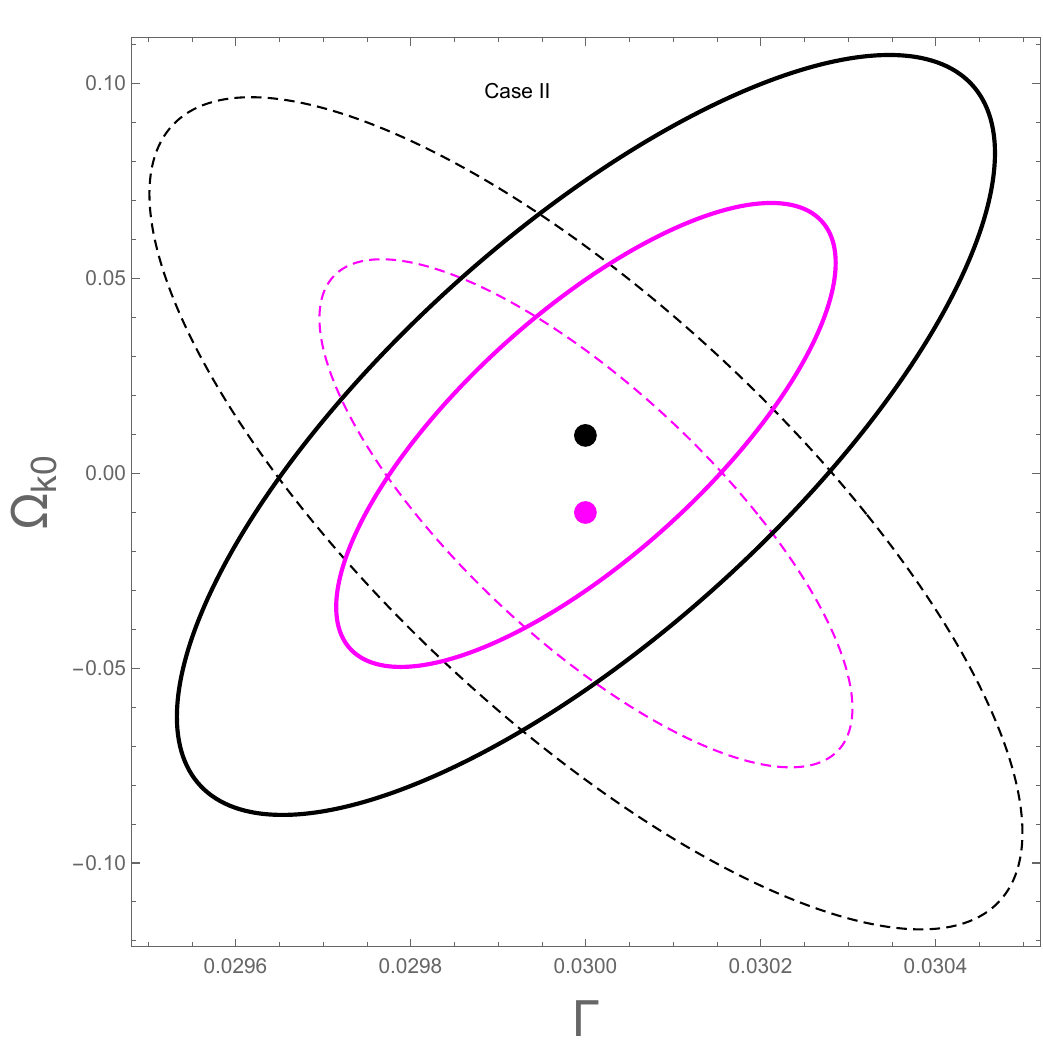}
\includegraphics[width=0.45\columnwidth]
{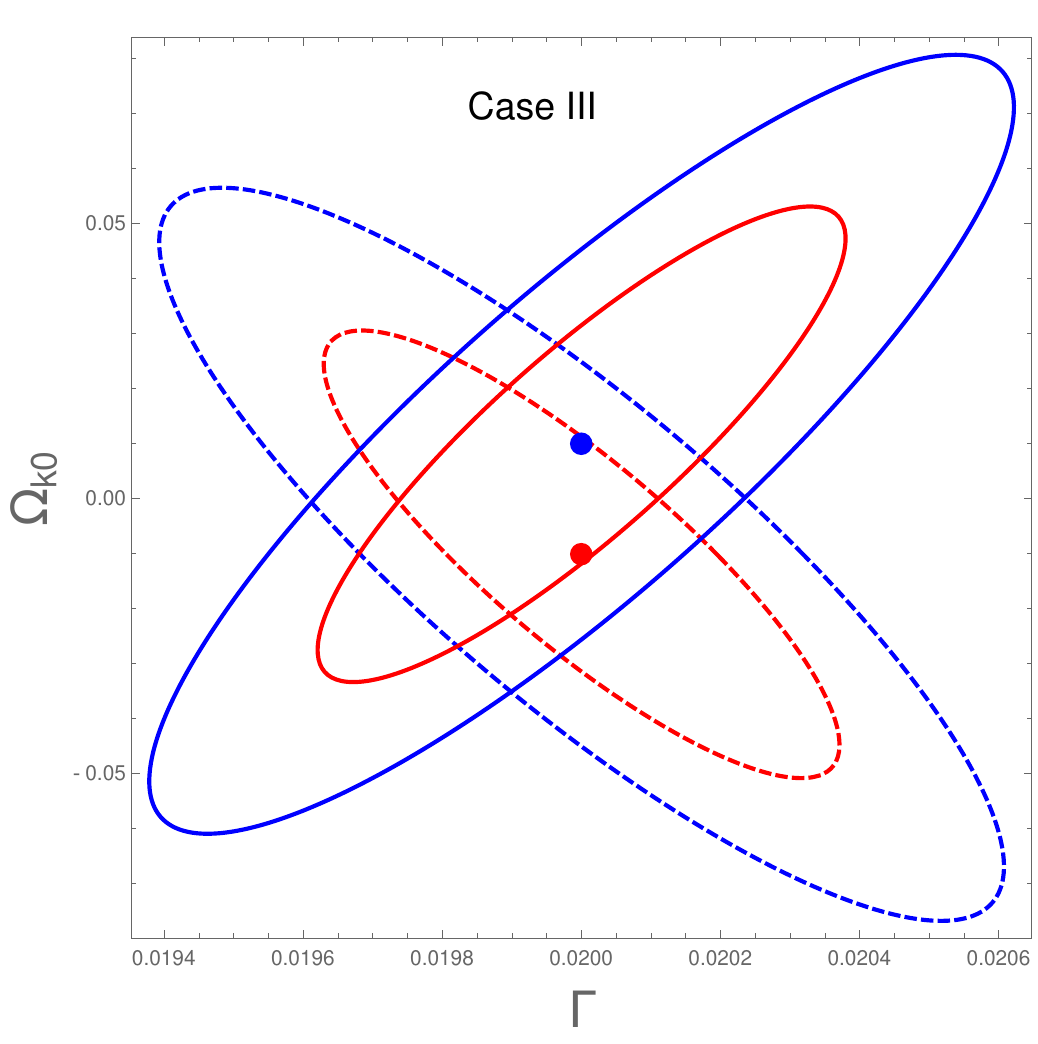}
\caption{\em 
The $1\sigma$ and $2\sigma$ confidence contours in $\Omega_{k0} - \Gamma$ parameter space for all three cases corresponding to CC + Pantheon dataset.}
\label{CCSN_H_rhode_fig}
\end{center}
\end{figure}
%%%%%%%%%%%%%%%%%%%%%%%%%%%%%%%%%%%%%
 As we can see from the plots, although the best-fit values of $\Omega_{k0}$ are small, comparatively larger values of $\Omega_{k0}$ are allowed at $1\sigma$ or $2\sigma$ confidence level. Furthermore, for positive curvature, $\Omega_{k0}$ is more tightly constrained as compared to the negative curvature case for all the datasets. It has also been found that for case I ($Q= - \Gamma H \rho_{de}$), Pantheon data puts very tight constraints on $\Gamma$ and $\Omega_{k0}$ showing slight preference for a spatially flat universe.
 
 In order to constrain the Barrow exponent $\Delta$ from observational data, we further present the $1\sigma$ and $2\sigma$ confidence contours in the $\Gamma - \Delta$ parameter space for CC, Pantheon and CC + Pantheon datasets corresponding to the three different interacting forms considered in this work. For this analysis, we have set $\Omega_{k0} \approx \pm 0.01$ for positive and negative curvature cases respectively since this happens to be the best-fit value for all the datasets as listed in table \ref{table1}. Figure \ref{CC_Delta_H_rhode_fig}, figure \ref{Pan_Delta_H_rhode_fig} and figure \ref{CCPan_Delta_H_rhode_fig} represent the $1\sigma$ and $2\sigma$ confidence contours in the $\Gamma - \Delta$ parameter space for CC, Pantheon and CC + Pantheon datasets respectively where case I, case II and case III refers to the same interacting forms mentioned earlier and the same colour coding and contour profiles have been used for all these plots. The corresponding best-fit values for the parameters $\Gamma$ and $\Delta$ 
for the different datasets has been listed in table \ref{table2}. As evident from table \ref{table2}, the best-fit values for the Barrow exponent $\Delta$ as well as the strength of interaction $\Gamma$ is very small.
As evident from the $\Gamma- \Delta$ plots in figure \ref{CC_Delta_H_rhode_fig}, we can see that for case I and case III, the cosmic chronometer data puts much tighter constraints on the parameters $\Delta$ and $\Gamma$ for positive curvature case (solid contour lines) as compared to negative curvature case (dotted contour lines). Also the best-fit values of the Barrow exponent $\Delta$ and the strength of interaction $\Gamma$ comes out to be very small even at the $1\sigma$ and $2\sigma$ confidence levels. This indicates that the quantum gravitational deformation effects are not very large for the proposed interacting Barrow holographic dark energy model which is consistent with earlier findings \cite{Luciano_2022, Jusufi:2021fek}. Again from figure \ref{Pan_Delta_H_rhode_fig} it is evident that for case II and case III, the constraints on $\Delta$ and $\Gamma$ are much tighter for positive curvature case. For negative curvature (see case III of figure \ref{Pan_Delta_H_rhode_fig}), though the best-fit value of $\Delta$ is small, $\Delta$ is not tightly constrained at $2\sigma$ level and can take any value up to $\Delta =1$. As seen from figure \ref{CCPan_Delta_H_rhode_fig}, the combined CC + Pantheon datasets put much tighter constraints on the parameters $\Delta$ and $\Gamma$ for all the three interacting models considered in this work.
 %%%%%%%%%%%%%%%%%%%%%%%%%%%%%%%%%%%%%
\begin{figure}[!h]
\begin{center}
\includegraphics[width=0.45\columnwidth]
{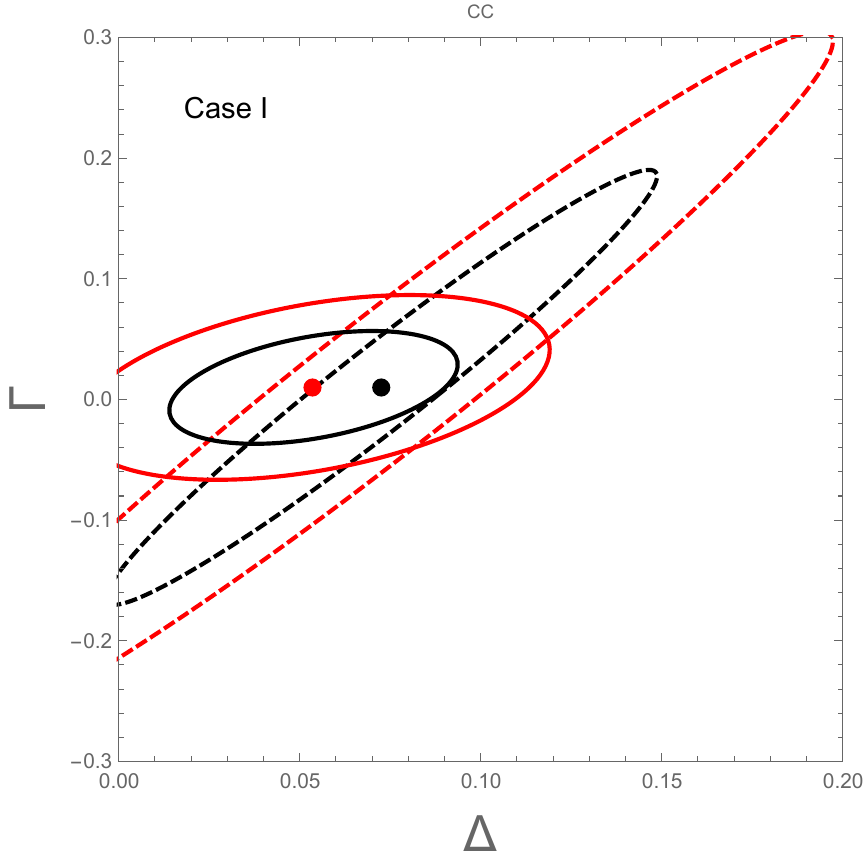} 
\includegraphics[width=0.45\columnwidth]
{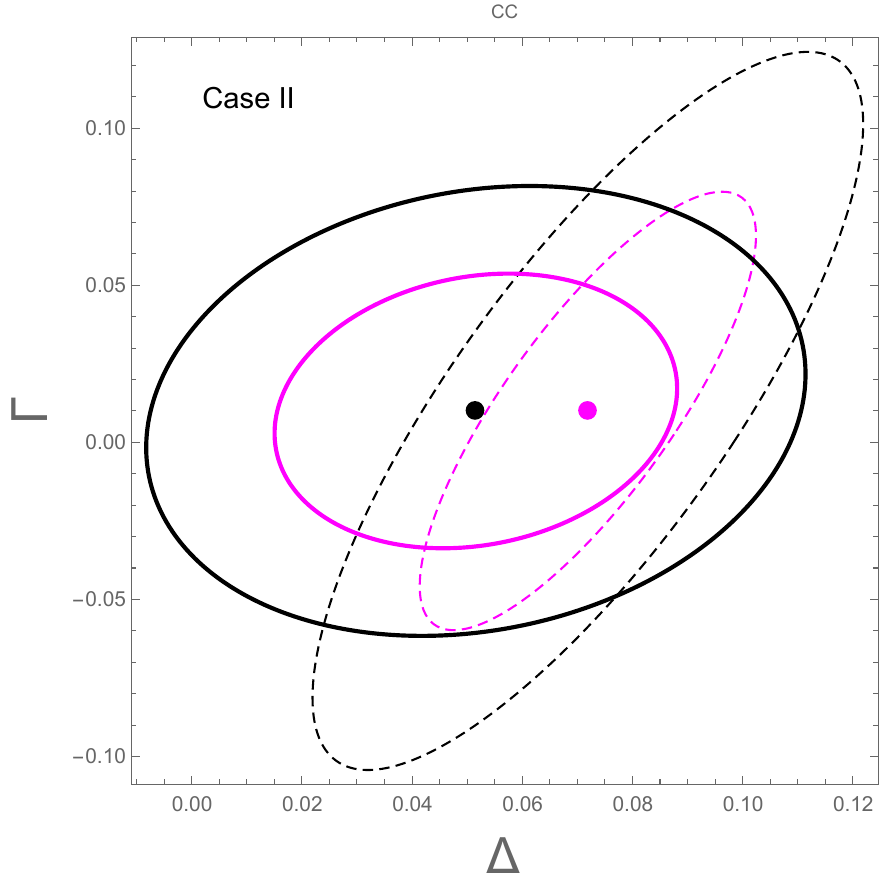}
\includegraphics[width=0.45\columnwidth]
{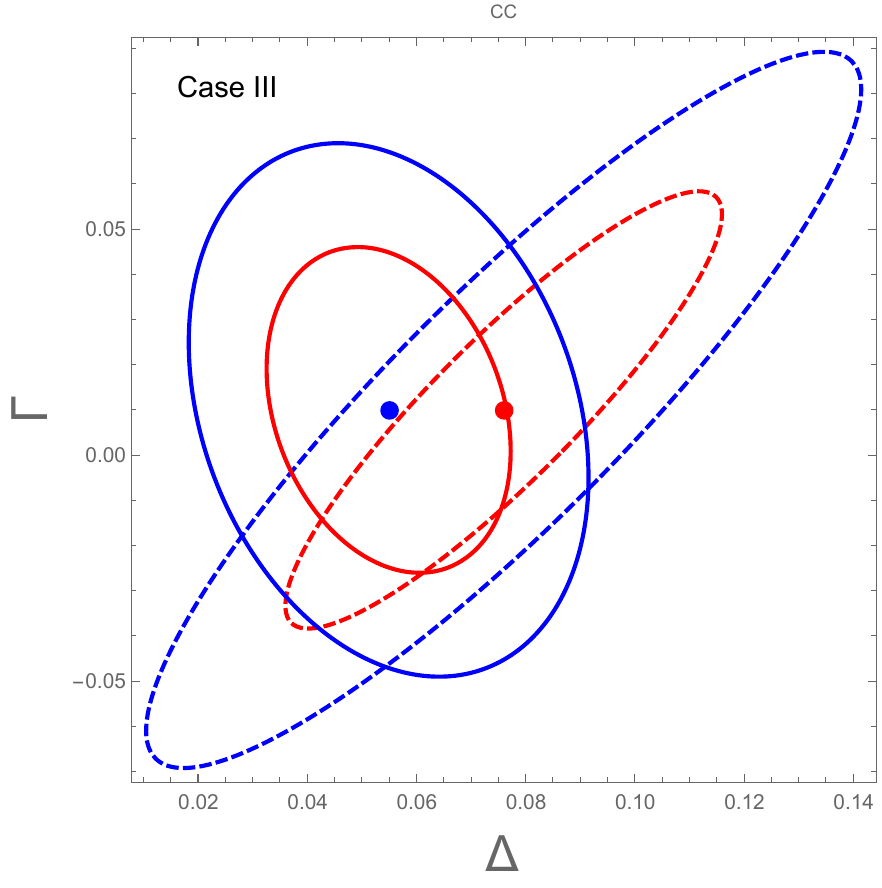}
\caption{\em 
The $1\sigma$ and $2\sigma$ confidence contours for all three cases in $\Gamma - \Delta$ parameter space corresponding to CC dataset. For all the panels, the solid lines correspond to the $1\sigma$ and $2\sigma$ confidence contours for the positive curvature case and the dashed lines correspond to the same for the negative curvature case.}
\label{CC_Delta_H_rhode_fig}
\end{center}
\end{figure}
%%%%%%%%%%%%%%%%%%%%%%%%%%%%%%%%%%%%%
 \begin{figure}[!h]
\begin{center}
\includegraphics[width=0.45\columnwidth]
{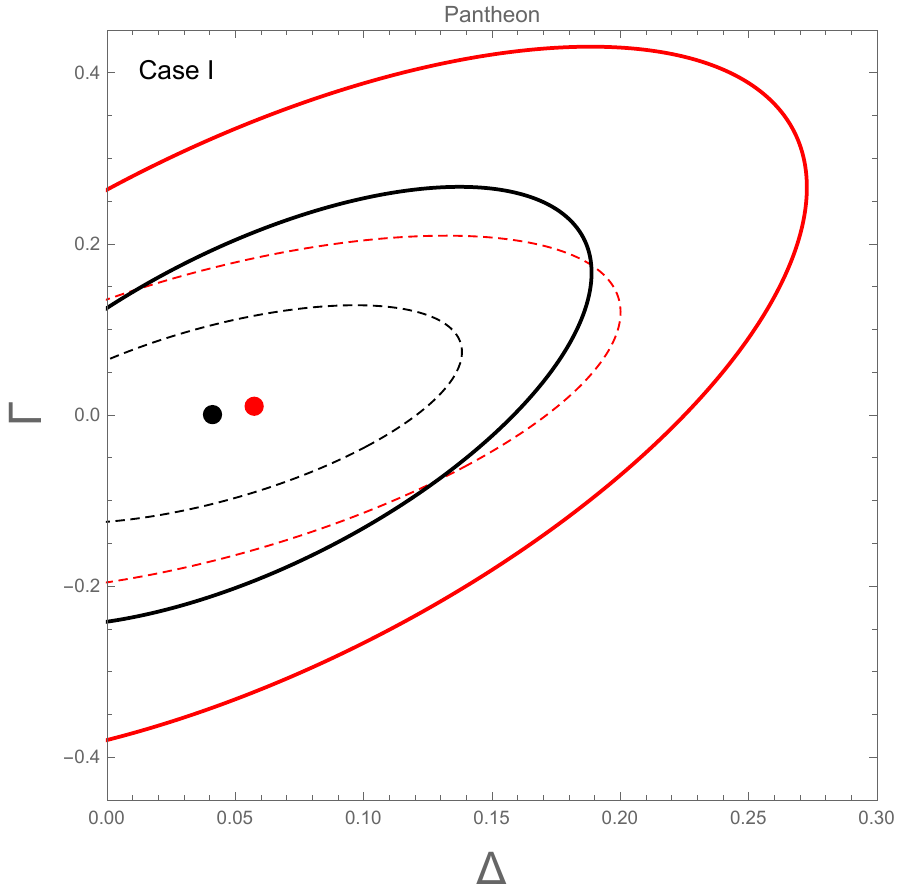} 
\includegraphics[width=0.45\columnwidth]
{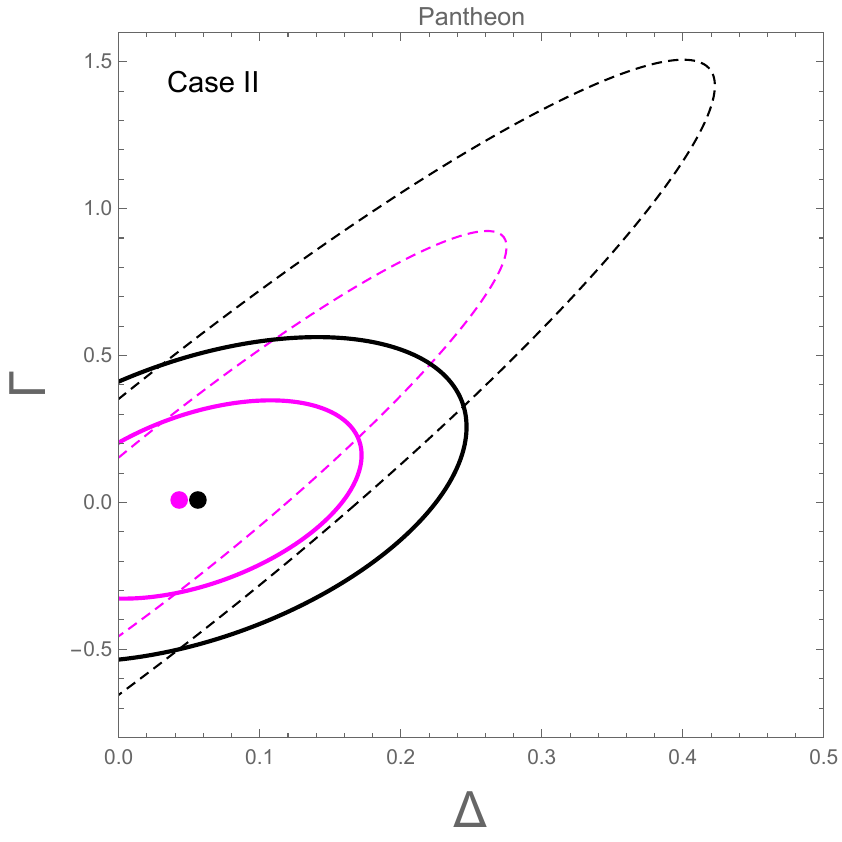}
\includegraphics[width=0.45\columnwidth]
{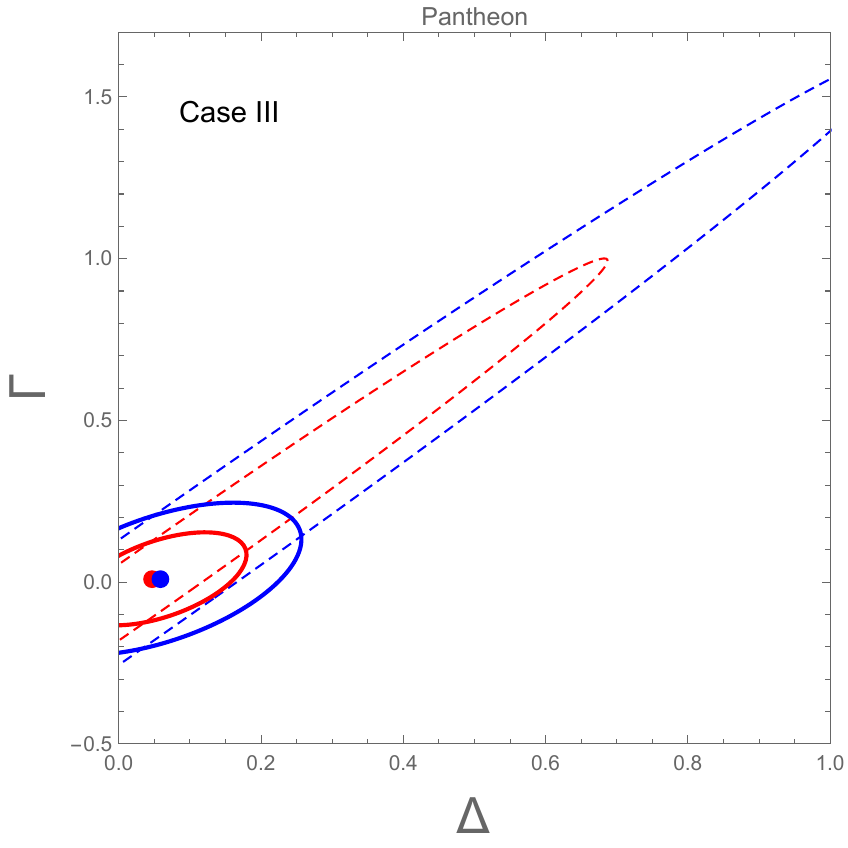}
\caption{\em 
The $1\sigma$ and $2\sigma$ confidence contours for all three cases in $\Gamma - \Delta$ parameter space corresponding to Pantheon dataset.}
\label{Pan_Delta_H_rhode_fig}
\end{center}
\end{figure}
 %%%%%%%%%%%%%%%%%%%%%%%%%%%%%%%%%%%%%
\begin{figure}[!h]
\begin{center}
\includegraphics[width=0.45\columnwidth]
{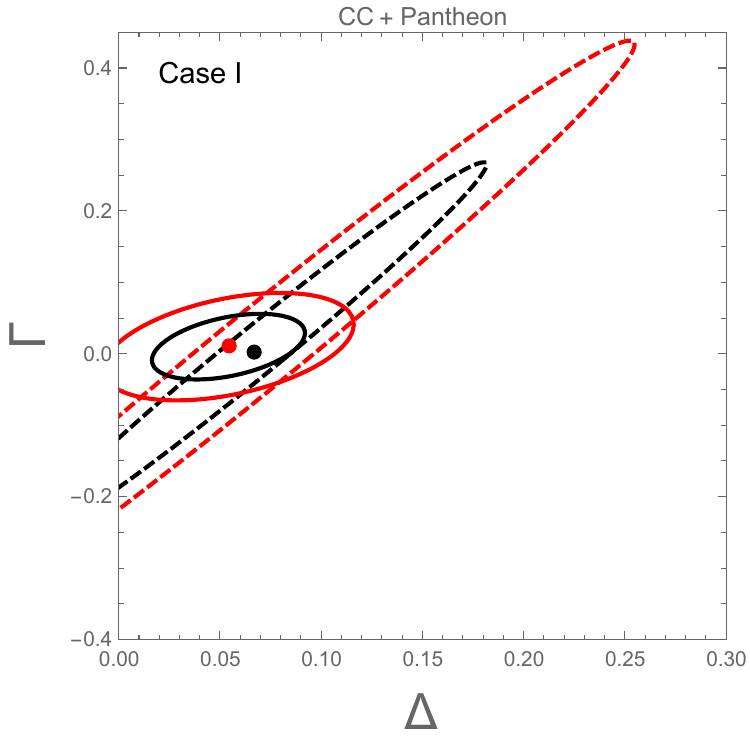} 
\includegraphics[width=0.45\columnwidth]
{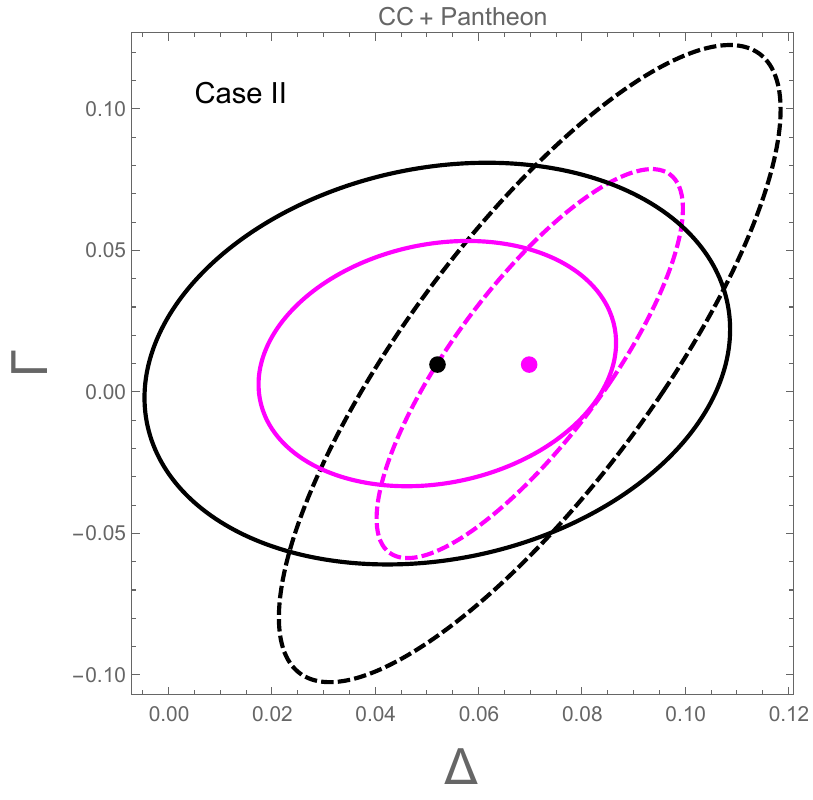}
\includegraphics[width=0.45\columnwidth]
{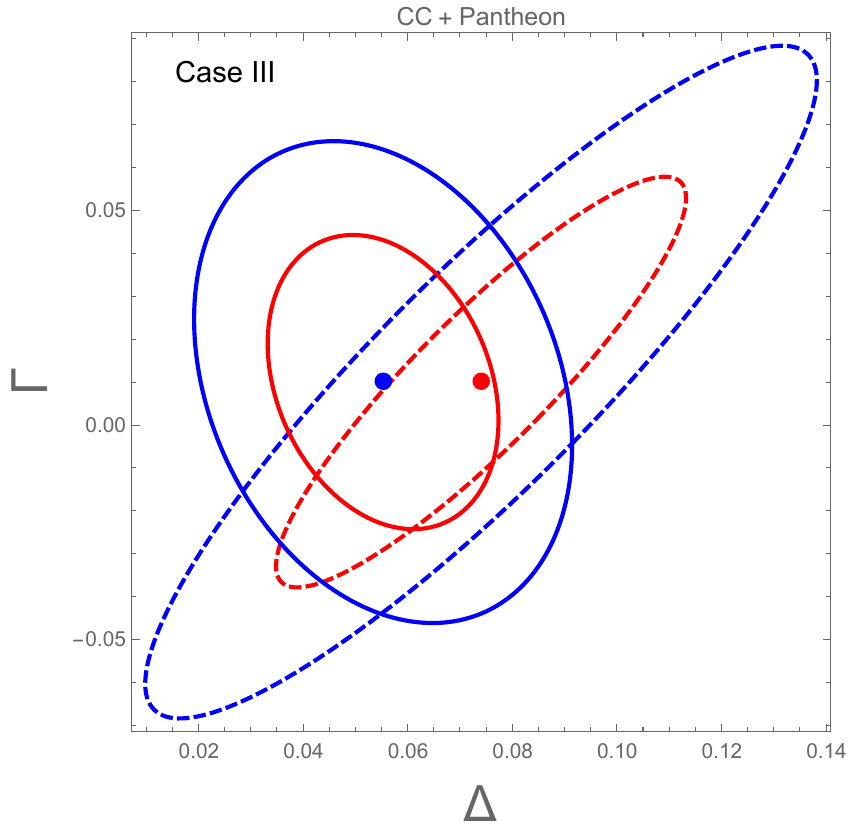}
\caption{\em 
The $1\sigma$ and $2\sigma$ confidence contours for all three cases in $\Gamma - \Delta$ parameter space  corresponding to CC + Pantheon dataset.}
\label{CCPan_Delta_H_rhode_fig}
\end{center}
\end{figure}
%%%%%%%%%%%%%%%%%%%%%%%%%%%%%%%%%%%%%%%
\begin{table}
\begin{center}
\begin{tabular}{ |c|c|c|c|c|c| } 
\hline
\hline
Model & Dataset & $k$ & $\Gamma$ & $\Delta$ & $\chi^2_{\text{min}}$ \\
\hline
\multirow{6}{4em}{Case I} 
&$CC$ & positive & $0.01$ & $0.053$ & $47.74$ \\ 
&& negative & $0.01$ & $0.073$ & $47.32$ \\
\cline{2-6}
&$Pantheon$ & positive & $0.01$ & $0.057$ & $34.38$ \\ 
&& negative & $0.001$ & $0.041$ & $33.88$ \\
\cline{2-6}
&$CC+Pantheon$ & positive & $0.01$ & $0.054$ & $82.13$ \\ 
&& negative & $0.001$ & $0.067$ & $81.06$ \\
\hline
\multirow{6}{4em}{Case II} 
&$CC$ & positive & $0.01$ & $0.052$ & $47.94$ \\ 
&& negative & $0.01$ & $0.072$ & $47.80$ \\
\cline{2-6}
&$Pantheon$ & positive & $0.01$ & $0.056$ & $34.37$ \\ 
&& negative & $0.01$ & $0.043$ & $33.91$ \\
\cline{2-6}
&$CC+Pantheon$ & positive & $0.01$ & $0.052$ & $82.31$ \\ 
&& negative & $0.01$ & $0.069$ & $82.06$ \\
\hline
\multirow{6}{4em}{Case III} 
&$CC$ & positive & $0.01$ & $0.055$ & $49.43$ \\ 
&& negative & $0.01$ & $0.076$ & $48.33$ \\
\cline{2-6}
&$Pantheon$ & positive & $0.01$ & $0.059$ & $34.47$ \\ 
&& negative & $0.01$ & $0.048$ & $33.95$ \\
\cline{2-6}
&$CC+Pantheon$ & positive & $0.01$ & $0.055$ & $83.91$ \\ 
&& negative & $0.01$ & $0.074$ & $82.64$ \\
\hline
\hline
\end{tabular}
\end{center}
\caption{Best-fit values of $\Gamma$ and $\Delta$ for interacting Barrow holographic dark energy in the case of non-flat universe for various datasets} \label{table2}
\end{table}
%%%%%%%%%%%%%%%%%%%%%%%%%%%%%%%%%%%
\begin{figure}[!h]
\begin{center}
\includegraphics[width=0.7\columnwidth]{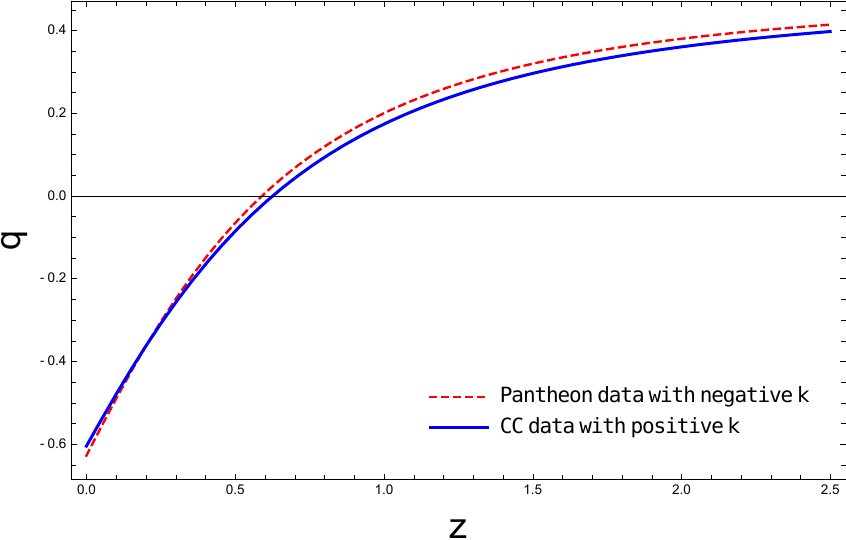} 
\caption{\em 
The evolution of deceleration parameter $q(z)$ for CC and Pantheon dataset}
\label{q-fig}
\end{center}
\end{figure}
%%%%%%%%%%%%%%%%%%%%%%%%%%%%%%%%%%%%%
 
 For the sake of completeness, we have also plotted the deceleration parameter $q(z)$ for the proposed interacting Barrow holographic dark energy model. In figure \ref{q-fig}, we have shown the evolution of the deceleration parameter considering the best-fit values for the CC dataset with positive curvature for case I and Pantheon dataset with the negative negative curvature for case II. However for all the best-fit values listed in table \ref{table1} corresponding to different datasets, the deceleration parameter $q(z)$ has been found to undergo a smooth transition from a decelerating to an accelerating phase, which depict the correct thermal history of the universe, consistent with the sequence of matter and dark energy dominated epochs. The transition redshift $z_t$ is found to vary slightly for different datasets, but is close to the widely accepted value of $z_t \sim 0.5$ \cite{riess2001farthest, Muthukrishna_2016}.  
\subsection{Statistical comparison of the model with the $\Lambda$CDM model}
For a comprehensive analysis, we perform a statistical comparison of the proposed interacting Barrow holographic dark energy model with the standard $\Lambda$CDM model. For a given set of competing models, the compatibility and relative tension between the models can be determined utilizing the statistical tools Akaike Information Criterion (AIC), Bayesian Information Criterion (BIC) and Deviance information criterion (DIC)  \cite{Liddle_2007, Spiegelhalter2002}. The AIC, BIC and DIC estimators are calculated from the maximum log-likelihood $\left({\mathcal{L}}_{max} = exp\left(\frac{-\chi^2}{2}\right)\right)$ of the model and is given by \cite{ModAIC, Goswami:2024ymh, Sol_Peracaula_2023}
\begin{eqnarray}
    AIC = -2~\mathrm{ln}~({\mathcal{L}}_{max}) + 2k  \\
BIC = -2~\mathrm{ln}~({\mathcal{L}}_{max}) + k~\mathrm{log}(N)\\
DIC =\chi^2(\bar{\theta}) +2 p_D
\end{eqnarray}
where $N$ is the count of data points utilized for
the sampling, $k$ is the number of independent model parameters, $\chi^2(\bar{\theta})$ represents the mean value of the $\chi^2$ function and $p_D = {\overline{\chi^2(\theta)}} - \chi^2(\bar{\theta})$ represents the effective number of parameters and is also called {\it model complexity} \cite{Liddle_2007}. 
Increase in the number of effective parameters in a model will generally lead to a better fit to the data (i.e., a lower $\chi^2(\bar{\theta})$), but this can lead to overfitting. In DIC, the models are penalized both by the value of $\chi^2(\bar{\theta})$, which favors a good fit, and by the effective number of parameters $p_D$. This prevents overfitting by ensuring that models with too many parameters are not favored purely due to their flexibility in fitting the data.

For comparison with the $\Lambda$CDM model, the relative difference
$\Delta AIC = |(AIC_{model} - AIC_{\Lambda CDM})|$, $\Delta BIC = |(BIC_{model} - BIC_{\Lambda CDM})|$ and $\Delta DIC = |(DIC_{model} - DIC_{\Lambda CDM})|$ are considered. According to Jeffrey’s scale \cite{Kass1995BayesF}, if $\Delta IC \le 2$ this will indicate that the model is strongly favoured; if it falls in the range $2 < \Delta IC < 6$, this indicates a mild tension between the models compared; if $\Delta IC \ge 10$, it indicates no favoured evidence and corresponds to a strong tension \cite{Anagnostopoulos:2020ctz}. In this work, we have compared our proposed model (considering the $\Gamma - \Delta$ parameter space) with the $\Lambda$CDM model and the relative $\Delta AIC$, $\Delta BIC$ and $\Delta DIC$ values have been listed in table \ref{tab:AICBIC}.
\begin{table}[h!]
\centering
\begin{tabular}{|c|c|c|c|c|c|c|c|c|}
 \hline
  \hline
Model & Dataset & Curvature &AIC &$\Delta$AIC  & BIC & $\Delta$BIC &DIC & $\Delta$DIC\\ 
\hline
$\multirow{9}{*}{Case I}$ &$\multirow{3}{*}{CC}$& $+1$ & $51.74$ & $1.64$ & $51.25$ & $1.87$ & $50.85$ & $1.79$\\
&  & $-1$& $51.32$ & $1.22$ & $50.83$ & $1.45$ & $50.23$ & $1.24$\\
& ($N=57$)& $\Lambda$CDM & $50.10$ & $0$ & $49.38$ & $0$ & $49.06$ & $0$\\
\cline{2-9}
  &$\multirow{3}{*}{Pantheon}$ & $+1$ & $38.38$ & $-1.39$ & $40.42$  & $-2.41$ & $37.93$ & $-1.18$ \\
  & & $-1$ & $37.88$  & $-1.89$ &  $39.92$ & $-2.91$ & $37.03$ & $-2.08$ \\
  & ($N=1048$)& $\Lambda$CDM & $39.77$  & $0$ &  $42.83$ & $0$ & $39.11$ & $0$ \\
  \cline{2-9}
  &$\multirow{3}{*}{CC + Pantheon}$ & $+1$ & $86.13$ & $1.59$ & $88.21$  & $0.55$ & $85.59$ & $1.9$ \\
  & & $-1$ & $85.06$  & $0.52$ &  $87.14$ & $-0.52$ & $84.93$ & $1.24$\\
  & ($N=1105$) & $\Lambda$CDM & $84.54$  & $0$ &  $87.66$ & $0$ & $83.69$ & $0$ \\
 \hline
   \hline
   $\multirow{9}{*}{Case II}$ &$\multirow{3}{*}{CC}$& $+1$ & $51.94$ & $1.84$ & $51.45$ & $2.07$ & $51.26$ & $2.2$\\
& & $-1$& $51.80$ & $1.70$ & $51.31$ & $1.93$ & $51.02$ & $1.96$\\
& ($N=57$) & $\Lambda$CDM & $50.10$ & $0$ & $49.38$ & $0$ & $49.06$ & $0$\\
\cline{2-9}
  &$\multirow{3}{*}{Pantheon}$ & $+1$ & $38.37$ & $-1.40$ & $40.41$  & $-2.42$ & $37.91$ & $-1.20$\\
  & & $-1 $& $37.91$  & $-1.86$ &  $39.95$ & $-2.88$ & $37.14$ & $-1.97$ \\
  & ($N=1048$) & $\Lambda$CDM & $39.77$  & $0$ &  $42.83$ & $0$ & $39.11$  & $0$  \\
  \cline{2-9}
  &$\multirow{3}{*}{CC + Pantheon}$ & $+1$ & $86.31$ & $1.77$ & $88.39$  & $0.73$ & $85.19$ & $1.5$\\
  & & $-1$ & $86.06$  & $1.52$ &  $88.14$ & $0.48$ & $85.02$ & $1.33$ \\
  &  ($N=1105$)& $\Lambda$CDM & $84.54$  & $0$ &  $87.66$ & $0$ & $83.69$ & $0$ \\
 \hline
 \hline
$\multirow{9}{*}{Case III}$ &$\multirow{3}{*}{CC}$& $+1$ & $53.43$ & $3.33$ & $52.94$ & $3.56$ & $52.13$ & $3.07$ \\
& & $-1$ & $52.33$ & $2.23$ & $51.84$ & $2.46$ & $50.37$ & $1.31$\\
& ($N=57$) & $\Lambda$CDM & $50.10$ & $0$ & $49.38$ & $0$ & $49.06$ & $0$\\
\cline{2-9}
  &$\multirow{3}{*}{Pantheon}$ & $+1$ & $38.47$ & $-1.30$ & $40.51$  & $-2.32$ & $37.63$ & $-1.48$\\
  & & $-1$ & $37.95$  & $-1.82$ &  $39.99$ & $-2.84$ & $37.10$ & $-2.01$ \\
  & ($N=1048$) & $\Lambda$CDM & $39.77$  & $0$ &  $42.83$ & $0$ & $39.11$ & $0$ \\
  \cline{2-9}
  &$\multirow{3}{*}{CC + Pantheon}$ & $+1$ & $87.91$ & $3.37$ & $89.99$  & $2.33$ & $86.73$ & $3.04$\\
  & & $-1$ & $86.64$  & $2.10$ &  $88.72$ & $1.06$ & $85.42$ & $1.73$\\
  & ($N=1105$) & $\Lambda$CDM & $84.54$  & $0$ &  $87.66$ & $0$ & $83.69$ & $0$  \\
 \hline
 \hline
\end{tabular}
\caption{The $AIC$, $BIC$, $DIC$ values for the proposed model in comparison with $\Lambda$CDM model and the corresponding relative differences $\Delta AIC$, $\Delta BIC$ and $\Delta DIC$}
\label{tab:AICBIC}
\end{table}
As noted from table \ref{tab:AICBIC}, the $\Delta IC$ values indicate that the proposed non-flat interacting Barrow holographic dark energy model is strongly favored in comparison to the standard $\Lambda$CDM model for the CC dataset. But for Pantheon datasets, $\Delta AIC$, $\Delta BIC$ values being negative indicates that $\Lambda$CDM model is preferred as compared to the proposed model for all the three cases. For CC + Pantheon datasets, it has been found that Case I and Case II models are strongly favoured as compared to $\Lambda$CDM model, but for case III, it exhibits mild tension.

\section{Conclusions}\label{conclusion}
In this work we have constructed an interacting model of Barrow holographic dark energy considering a non-flat spatial curvature of the universe. Barrow holographic dark energy (BHDE) models provide an alternative approach to understanding cosmic acceleration and the evolution of the universe. These models offer an intriguing perspective on addressing the nature of dark energy through the lens of holographic principles and Barrow entropy modifications. In this work we have studied the cosmological implications of BHDE model considering interaction between the holographic dark energy component and the matter component. We have phenomenologically chosen three different forms of interaction which happen to be popular choices for the interaction term. Considering closed and open spatial geometry we have obtained evolution equations for energy densities of the dark energy and dark matter components and have also obtained analytical expressions for the equation of state parameter $w_{DE}$ for each case. 
The motivation of this work was to explore the cosmological implications of an interaction between Barrow holographic dark energy and dark matter components and to check whether interacting BHDE models are supported by the current cosmological observations. Our analysis shows that the interacting BHDE model can account for the observed accelerated expansion of the universe with a variable equation of state that evolves over time. This adaptability allows the model to accommodate various cosmological scenarios.\\ 
We have looked into the effect of the Barrow exponent $\Delta$, the strength of interaction $\Gamma$ as well as the curvature density parameter $\Omega_{k0}$ on the equation of state parameter $w_{DE}(z)$. It has been observed that for both open and closed curvature cases, $w_{DE}$ lies in the phantom region at present. For $\Delta =0$, the equation of state parameter for the IBHDE lies completely in the quintessence regime, but for higher values of $\Delta$, a phantom-divide crossing is observed in the past. This is consistent with the results obtained earlier \cite{Adhikary:2021}. Similarly from figure \ref{model1diffgamma} it has been found that when we set $\Delta=0.2$, the universe was lying mostly in the phantom regime. For different values of the strength of interaction, the redshift for phantom divide crossing is shifted slightly for both open and closed cases. Also the closed universe models are found to enter the phantom regime earlier as compared
to open model for all the phenomenologically chosen interaction terms. This findings contradicts the results obtained by Kim et al. \cite{Kim_2006} which claims that an interacting holographic dark energy model cannot accommodate a transition from the dark energy with $w_{eff}^{\Lambda} \ge -1$ to the phantom regime with $w_{eff}^{\Lambda} < -1$. This work shows that one can generate a phantom-like mixture from an interaction between the Barrow holographic dark energy component and the dark matter component. We have also provided a comparison of the proposed non-flat interacting Barrow holographic dark energy with the concordance $\Lambda$CDM paradigm using the AIC, BIC and DIC information criteria. It has been found that the proposed model is statistically compatible with $\Lambda$CDM model for CC dataset. But for CC + Pantheon dataset, it exhibits mild tension. Further investigations are required to refine the various parameters of the model and determine its compatibility with other cosmological phenomena, such as structure formation and early-universe behavior. Furthermore, testing the Barrow holographic dark energy models, non-interacting or interacting, alongside other dark energy models may deepen our understanding of the subtleties within dark energy physics and the role of quantum effects in cosmic evolution.
\section{Acknowledgement}
This work is supported by Anusandhan National Research Foundation (ANRF), Government of India through the project CRG/2023/000185. SD would also like to acknowledge IUCAA, Pune for providing support through associateship program. The authors would like to acknowledge the support and facilities under ICARD, Pune at Department of Physics, Visva-Bharati, Santiniketan.   

\appendix
\section{Detailed calculations for $k=+1$ case}\label{appendix1}
Continuity equation for interacting barrow holographic dark energy is given by,
\begin{equation}
{\dot{\rho}}_{DE} + 3 H \left( 1+w_{DE}\right)\rho_{DE} = -Q=\Gamma H \rho_{DE}\\
\end{equation}
\begin{equation}
\implies  \int \frac{d\rho_{DE}}{\rho_{DE}} = \int \left[\Gamma H  - 3 H \left( 1+w_{DE}\right) \right] \,dt\nonumber\\
\end{equation}
\begin{equation}
\implies  \ln(\rho_{DE}) = \int \left[\Gamma H  - 3 H \left( 1+w_{DE}\right) \right] \,dt + \ln(\rho_{0})\nonumber\\
\end{equation}
\begin{equation}
\implies  \ln\left(\frac{\rho_{DE}}{\rho_{0}}\right) = \int \left[\Gamma H  - 3 H \left( 1+w_{DE}\right) \right] \,dt \nonumber\\
\end{equation}
\begin{equation}
\implies \frac{d}{dx}\left[ \ln\left(\frac{\rho_{DE}}{\rho_{0}}\right)\right] = \frac{d}{dx}\left[\int \left[\Gamma H  - 3 H \left( 1+w_{DE}\right) \right]\right] \,dt \nonumber\\
\end{equation}
\begin{equation}\label{ddx}
\implies \frac{d}{dx}\left[ \ln\left(\frac{\rho_{DE}}{\rho_{0}}\right)\right] =  \left[\Gamma   - 3  \left( 1+w_{DE}\right) \right] 
\end{equation}
Again L.H.S of equation \ref{ddx} we can write as,
\begin{eqnarray}
\frac{d}{dx}\left[ \ln\left(\frac{\rho_{DE}}{\rho_{0}}\right)\right] &=&  \frac{1}{\rho_{DE}/\rho_{0}} \frac{d}{dx}\left(\frac{\rho_{DE}}{\rho_{0}}\right)\nonumber\\
 &=& \frac{1}{\rho_{DE}} \frac{d}{dx} \left(\Omega_{DE} \rho_{cr}\right)\nonumber\\
&=& \frac{1}{\rho_{DE}} \frac{d}{dx} \left(\frac{\Omega_{DE}}{\left(1-\Omega_{DE}\right)} \left(1-\Omega_{DE}\right)\rho_{cr}\right)\nonumber\\
&=& \frac{1}{\rho_{DE}} \frac{d}{dx} \left(\frac{\Omega_{DE}}{\left(1-\Omega_{DE}\right)} \left(\rho_{m}-\frac{\beta}{a^2}\right)\right)\nonumber\\
&=& \left[\frac{\Omega_{DE}^\prime}{\Omega_{DE}\left(1-\Omega_{DE}\right)} + \frac{1}{\left(\rho_{m}-\frac{\beta}{a^2}\right)}\frac{d}{dx}\left(\rho_{m}-\frac{\beta}{a^2}\right)\right]\nonumber\\
\end{eqnarray}
Now continuity equation for matter density can be written as,
\begin{eqnarray}\label{ddxrhom}
{\dot{\rho}}_{m} + 3 H \rho_{m} = Q =-\Gamma H \rho_{DE}\nonumber\\
\implies \int \frac{d\rho_{m}}{\rho_{m}}= - \int \left( \frac{\Gamma H}{r} +3 H\right) \,dt \nonumber \\
\implies  \ln(\rho_{m}) = - \int \left( \frac{\Gamma H}{r} +3 H\right) \,dt + \ln(\rho_{1})\nonumber\\
\implies  \ln\left(\frac{\rho_{m}}{\rho_{1}}\right) = - \int \left( \frac{\Gamma H}{r} +3 H\right) \,dt \nonumber\\
\implies \frac{d}{dx} \left[ \ln\left(\frac{\rho_{m}}{\rho_{1}}\right)\right] = - \frac{d}{dx}\left[ \int \left( \frac{\Gamma H}{r} +3 H\right) \,dt \right]\nonumber\\
\implies \frac{1}{\rho_{m}}\frac{d}{dx}\left(\rho_{m}\right) = - \left[   \frac{\Gamma }{r} +3 \right]
\end{eqnarray}
Using equation (\ref{ddxrhom}) we can write,
\begin{eqnarray}
\frac{1}{\left(\rho_{m}-\frac{\beta}{a^2}\right)}\frac{d}{dx}\left(\rho_{m}-\frac{\beta}{a^2}\right) = \frac{1}{1-\Omega_{DE}}\left[ -\Omega_{m}\left(\frac{\Gamma}{r}+3\right)+2\left(\Omega_{m}+\Omega_{DE}-1\right)\right]
\end{eqnarray}
Finally we have arrived at the following equations,
\begin{equation}
\begin{split}
\frac{\Omega_{DE}'}{\Omega_{DE} (1-\Omega_{DE})} = (\Delta - 2)\left(1-\cos y \sqrt{\frac{3 M_p^2 \Omega_{DE}}{C}} L^{-\frac{\Delta}{2}}\right)\\ - \frac{1}{(1-\Omega_{DE})} \left[ 2(\Omega_{m}+\Omega_{DE}-1)-\Omega_{m}\left(\frac{\Gamma}{r}+3\right)\right]
\end{split}
\end{equation}
and
\begin{equation}
\begin{split}
\frac{\Omega_{m}'}{\Omega_{m}} =-\left(\frac{\Gamma}{r}+3\right)-\Omega_{DE} (\Delta - 2)(1-\cos y \sqrt{\frac{3 M_p^2 \Omega_{DE}}{C}} L^{-\frac{\Delta}{2}})\\ - \left[ 2(\Omega_{m}+\Omega_{DE}-1)-\Omega_{m}\left(\frac{\Gamma}{r}+3\right)\right],
\end{split}
\end{equation}
\section{Detailed calculations for $k=-1$ case}\label{appendix2}
Continuity equation for interacting barrow holographic dark energy density,
\begin{eqnarray}\label{ddxn}
{\dot{\rho}}_{DE} + 3 H \left( 1+w_{DE}\right)\rho_{DE} = -Q=\Gamma H \rho_{DE} \nonumber\\
\implies  \int \frac{d\rho_{DE}}{\rho_{DE}} = \int \left[\Gamma H  - 3 H \left( 1+w_{DE}\right) \right] \,dt\nonumber\\
\implies  \ln(\rho_{DE}) = \int \left[\Gamma H  - 3 H \left( 1+w_{DE}\right) \right] \,dt + \ln(\rho_{0})\nonumber\\
\implies  \ln\left(\frac{\rho_{DE}}{\rho_{0}}\right) = \int \left[\Gamma H  - 3 H \left( 1+w_{DE}\right) \right] \,dt \nonumber\\
\implies \frac{d}{dx}\left[ \ln\left(\frac{\rho_{DE}}{\rho_{0}}\right)\right] = \frac{d}{dx}\left[\int \left[\Gamma H  - 3 H \left( 1+w_{DE}\right) \right]\right] \,dt \nonumber\\
\implies \frac{d}{dx}\left[ \ln\left(\frac{\rho_{DE}}{\rho_{0}}\right)\right] =  \left[\Gamma   - 3  \left( 1+w_{DE}\right) \right] 
\end{eqnarray}

Again  L.H.S of equation (\ref{ddxn}) we can write as,
\begin{eqnarray}
\frac{d}{dx}\left[ \ln\left(\frac{\rho_{DE}}{\rho_{0}}\right)\right] &=&  \frac{1}{\rho_{DE}/\rho_{0}} \frac{d}{dx}\left(\frac{\rho_{DE}}{\rho_{0}}\right)\nonumber\\
                         &=& \frac{1}{\rho_{DE}} \frac{d}{dx} \left(\Omega_{DE} \rho_{cr}\right)\nonumber\\
                        &=& \frac{1}{\rho_{DE}} \frac{d}{dx} \left(\frac{\Omega_{DE}}{\left(1-\Omega_{DE}\right)} \left(1-\Omega_{DE}\right)\rho_{cr}\right)\nonumber\\
                        &=& \frac{1}{\rho_{DE}} \frac{d}{dx} \left(\frac{\Omega_{DE}}{\left(1-\Omega_{DE}\right)} \left(\rho_{m}+\frac{\beta}{a^2}\right)\right)\nonumber\\
                        &=& \left[\frac{\Omega_{DE}^\prime}{\Omega_{DE}\left(1-\Omega_{DE}\right)} + \frac{1}{\left(\rho_{m}+\frac{\beta}{a^2}\right)}\frac{d}{dx}\left(\rho_{m}+\frac{\beta}{a^2}\right)\right]
\end{eqnarray}
Now continuity equation for matter density can be written as,
\begin{eqnarray}\label{ddxrhomn}
{\dot{\rho}}_{m} + 3 H \rho_{m} = Q =-\Gamma H \rho_{DE}\nonumber\\
\implies \int \frac{d\rho_{m}}{\rho_{m}}= - \int \left( \frac{\Gamma H}{r} +3 H\right) \,dt \nonumber \\
\implies  \ln(\rho_{m}) = - \int \left( \frac{\Gamma H}{r} +3 H\right) \,dt + \ln(\rho_{1})\nonumber\\
\implies  \ln\left(\frac{\rho_{m}}{\rho_{1}}\right) = - \int \left( \frac{\Gamma H}{r} +3 H\right) \,dt \nonumber\\
\implies \frac{d}{dx} \left[ \ln\left(\frac{\rho_{m}}{\rho_{1}}\right)\right] = - \frac{d}{dx}\left[ \int \left( \frac{\Gamma H}{r} +3 H\right) \,dt \right]\nonumber\\
\implies \frac{1}{\rho_{m}}\frac{d}{dx}\left(\rho_{m}\right) = - \left[   \frac{\Gamma }{r} +3 \right]
\end{eqnarray}
Using equation (\ref{ddxrhomn}) we can write,
\begin{eqnarray}
\frac{1}{\left(\rho_{m}+\frac{\beta}{a^2}\right)}\frac{d}{dx}\left(\rho_{m}+\frac{\beta}{a^2}\right) = \frac{1}{1-\Omega_{DE}}\left[ -\Omega_{m}\left(\frac{\Gamma}{r}+3\right)+2\left(\Omega_{m}+\Omega_{DE}-1\right)\right]
\end{eqnarray}

Finally we have reached the following equations for negative curvature universe,
\begin{equation}
\begin{split}
\frac{\Omega_{DE}'}{\Omega_{DE} (1-\Omega_{DE})} = (\Delta - 2)\left(1-\cosh y \sqrt{\frac{3 M_p^2 \Omega_{DE}}{C}} L^{-\frac{\Delta}{2}}\right)\\ - \frac{1}{(1-\Omega_{DE})} \left[ 2(\Omega_{m}+\Omega_{DE}-1)-\Omega_{m}\left(\frac{\Gamma}{r}+3\right)\right]
\end{split}
\end{equation}
and
\begin{equation}
\begin{split}
\frac{\Omega_{m}'}{\Omega_{m}} =-\left(\frac{\Gamma}{r}+3\right)-\Omega_{DE} (\Delta - 2)(1-\cosh y \sqrt{\frac{3 M_p^2 \Omega_{DE}}{C}} L^{-\frac{\Delta}{2}})\\ - \left[ 2(\Omega_{m}+\Omega_{DE}-1)-\Omega_{m}\left(\frac{\Gamma}{r}+3\right)\right],
\end{split}
\end{equation}

\bibliographystyle{JHEP}
\bibliography{myref.bib} 
\end{document}